\documentclass[journal]{IEEEtran}
\usepackage{amsmath,amsfonts,amsthm}
\usepackage{algorithmic}
\usepackage{algorithm}
\usepackage{array}
\usepackage{bm}
\usepackage[caption=false,font=normalsize,labelfont=sf,textfont=sf]{subfig}
\usepackage{textcomp}
\usepackage{url}
\usepackage{verbatim}
\usepackage{graphicx}
\usepackage{cite}
\usepackage{xcolor}
\usepackage{hyperref}
\hypersetup{hidelinks}
\newtheorem{theorem}{Theorem}
\newtheorem{lemma}[theorem]{Lemma}
\newtheorem{remark}[theorem]{Remark}

\newcommand{\Qgate}{ Q^\text{gate}}
\newcommand{\Qout}{ Q^\text{out}}
\newcommand{\Qin}{ Q^\text{in}}

\newcommand{\ds}{\mathrm{ds}}
\newcommand{\us}{\mathrm{us}}
\newcommand{\hrref}{ h^\text{ref}}
\newcommand{\hds}{ h^\text{ds}}
\newcommand{\Aref}{ A^\text{ref}}
\newcommand{\red}{\color{black}}
\newcommand{\blue}{\color{black}}
\newcommand\mc[1]{{\color{black} #1}}

\newcommand{\AQ}{ \left( \begin{matrix} \tilde{A}(x,t)\\ \tilde{Q}(x,t) \end{matrix}\right)}
\newcommand{\AQs}{ \left( \begin{matrix} \hat{A}(x,s)\\ \hat{Q}(x,s) \end{matrix}\right)}
\newcommand{\AQin}{ \left( \begin{matrix} \hat{A}(0,s)\\ \hat{Q}(0,s) \end{matrix}\right)}
\newcommand{\AQout}{ \left( \begin{matrix} \hat{A}(L,s)\\ \hat{Q}(L,s) \end{matrix}\right)}
\renewcommand{\i}{\mathrm{j}}

\begin{document}

\title{Decentralized water-level balancing for irrigation channels in storage critical operations\thanks{Funded in part by the Australian Research Council (Grant Number: LP200200917) and Rubicon Water, VIC 3123, Australia. Aspects of the work relate to International Patent Application (PCT/AU2025/051191).}% <-this % stops a space
}

\author{Timm Strecker and Michael Cantoni%~\IEEEmembership{Member,~IEEE}% <-this % stops a space
\thanks{T. Strecker is with the Department of Electrical and Electronic Engineering, The University of Melbourne, VIC 3010, Australia. (email: {\tt timm.strecker@unimelb.edu.au})}%
\thanks{M. Cantoni is with the Department of Electrical and Electronic Engineering, The University of Melbourne, VIC 3010, Australia. Corresponding author. (email: {\tt cantoni@unimelb.edu.au}) }%
}



\maketitle

\begin{abstract}
A feedback control \mc{system} is proposed for balancing the deviations of water levels from set-points along open channels subject to uncertain supply-demand mismatch that exceeds individual pool capacity. Decentralized controllers adjust \mc{the gate} flows between pools to regulate \mc{potentially weighted} differences between neighbouring water-level errors to zero in steady state. A sequential SISO loop-shaping procedure is developed for the design of each local flow controller based on distributed parameter transfer function models of the \mc{channel} dynamics. Recursive feasibility of the procedure for relevant performance specifications, and stability of the resulting MIMO closed-loop, are verified by supporting analysis. \mc{Both} numerical simulations \mc{and field trial results} are presented.
\end{abstract}


\section{Introduction}\label{sec:introduction}
Irrigation supports agriculture throughout the world \cite{UNESCO2019water}.  Often, large-scale networks of open channels are used to deliver water under the power of gravity from a source (e.g., reservoir or river) to irrigators. Flow throughout the network is controlled by adjustable gates located along the channels. The gates divide the channels into shorter sections called pools. In practice, the use of automatic control has led to major improvements over manual operation, in terms of both efficiency and quality of service; e.g., see~\cite{conde2021modeling,cantonimareels2021encyclopedia,wahlin2014canal,malaterre2007control,mareels2005systems}. The efficiency improvements relate to reduction of outflow losses at the ends of channels. The quality of service improvements stem from tighter regulation of the water levels at the off-take points that serve farms and channel branches. {\red Each of these water levels} reflects the local capacity for gravity fed supply of flow, \mc{as well as the volume of water stored, and thus, the available storage, in the corresponding pool.}

For a variety of reasons, including communication infrastructure limitations and matters pertaining to controller design and maintenance, the use of decentralized control is preferred at the gate level. In such a set up, each gate has a dedicated controller that can access measurements from the two pools immediately upstream and downstream only. Two common decentralized control architectures are upstream level control and distant downstream control \cite[Sec.~7.2]{litrico2009modeling}. 
With upstream level control, the flow over each gate is used to regulate the water level directly upstream of the gate, where off-takes are typically located. Such an operating mode is supply driven. Any mismatch between off-take demand and upstream supply is propagated to the next pool in the downstream direction, which can lead to losses in the case of over supply that exceeds the storage capacity of the most downstream pool, or service degradation downstream of the mismatch in the case of under supply. With distant downstream control~\cite{malaterre1998modeling,schuurmans1999simple,weyer2002decentralised,dehalleux2003boundary,litrico2003advanced,dulhoste2004nonlinear,litrico2005experimental}, each gate is used to control the water level at the distant end of the downstream pool, where off-takes are located. In this way, the levels are regulated via adjustment of pool inflows. This mode of operation results in demand driven release of flow from the source in order to match changes in outflows. Downstream storage is not used in the response, which makes it possible to set the flow at the end of the channel to zero, eliminating losses. However, distant downstream control relies on full authority over the inflow at the source to the channel. Both the upstream level and distant downstream control architectures involve directed information flow, downstream in the former, and upstream in the latter.

With distant downstream control, full authority is required over the flow at the source to the corresponding stretch of channel. There are situations in which this is not the case. For example, when the source flow can be adjusted just once or twice a day, compared to the timescale of minutes for gate adjustments under distant downstream control, or when other source limits are insufficient to service transient flow peaks. Situations where additional flow must be absorbed from the source as part of flood management are also relevant. In such cases, standard distant downstream control all the way to the source leads to loss of control over the most upstream pool. Given its limited capacity, the first pool could flood when there is over supply, or drain to a level where downstream flow cannot be delivered, when there is under supply.  

An alternative decentralized feedback control architecture is proposed here to achieve \mc{a} balanced spread of supply-demand mismatch among multiple pools along a stretch of channel \mc{under storage critical operating conditions}. Each gate flow is used to steer the difference between the water-level errors in the neighbouring upstream and downstream pools towards zero. By operating the most upstream pools of a channel in this way, the response to a constant volume mismatch between supply and demand is such that the water level deviations in the top pools converge to the same value. In particular, storage capacity of the top pools is effectively combined to reduce the impact of supply-demand mismatch, compared to the use of a single pool. Further downstream, the rest of the channel can operate in the distant downstream configuration, drawing water from the balanced pools, enabling a demand-driven mode of operation overall. Variations of the balancing objective are possible, which, e.g., prioritize tighter level control in some pools \mc{over} others.

Unlike upstream level and distant downstream control, the proposed decentralized balancing control system leads to bi-directional information flow; i.e., transients propagate spatially in both the upstream and downstream directions. Therefore, ensuring stability of the MIMO closed-loop dynamics requires some care. One possibility is to employ a stability constrained semi-definite programming based $\mathcal{H}_\infty$ distributed controller synthesis~\cite{dullerud2004distributed,langbort2004distributed}. This approach would involve the construction of a state-space model of the channel, and the design of weights that \mc{reflect} performance objectives, \mc{as} considered in~\cite{li2005water,cantoni2007control,li2008distributed} for distant downstream controller synthesis from the perspective of $\mathcal{H}_\infty$ loop-shaping~\cite{mcfarlane1992loop}. MIMO closed-loop stability in the distant downstream control case is a matter of achieving stability pool-by-pool because of the directed information flow. Thus, the design of decentralized compensators is straightforward \mc{at least in terms of achieving  stability}. \mc{In \cite{li2005water,cantoni2007control,li2008distributed},} the \mc{motivation for considering} MIMO $\mathcal{H}_\infty$ synthesis of a distributed augmentation of \mc{local} SISO loop-shaping compensators \mc{for each gate} stems from fundamental challenges associated with the spatial propagation of transient flow peaks, which is a \mc{channel level (i.e., global)} performance characteristic.

Since information flow is bi-directional in the case of balancing control, the design of useful local gate compensator weights for $\mathcal{H}_\infty$ loop-shaping distributed controller synthesis becomes more intricate. The sequential SISO \mc{classical} loop-shaping method proposed for the direct design of decentralized balancing gate controllers arose from investigating the systematic design such weights. Experience suggests that only marginal benefit is gained by augmenting these local compensators with the distributed controller resulting from $\mathcal{H}_\infty$ synthesis based refinement, despite considerable increase in the overall control system complexity. \mc{Therefore,} the focus henceforth is development of the sequential loop-shaping method and supporting analytical results. \mc{As the main technical contribution}, it is established that the sequential SISO procedure directly yields decentralized PI\mc{-type} gate controllers that achieve MIMO closed-loop stability and desirable performance without further augmentation. To the engineer, these simple controllers are preferred for deployment. 

The consensus control literature provides several stability certificates for systems with similar interconnection structures, including the distributed PI control of integrators in~\cite{andreasson2014distributed}, and PID control of networked first-order systems in~\cite{lombana2014distributedPID}. Related approaches to balancing in flow networks have been studied in, e.g., \cite{hui1999hydrodynamic,wei2013loadbalancing,burger2015dynamic}. However, the systems considered involve compartments that exhibit integrator and other forms of passive dynamics. By contrast, irrigation channel pools exhibit non-passive distributed parameter dynamics, including delay and other resonant hydraulic effects. The proposed sequential SISO loop-shaping method involves {\red corresponding} frequency domain transfer function models for the pool dynamics. These are derived via linearization of Saint Venant hyperbolic partial differential equation models.

The paper is organized as follows. The distributed parameter model of an irrigation channel is introduced in Section \ref{sec: modelling}. Section \ref{sec: control design} includes the \mc{proposed} control design \mc{method}, and supporting analysis to establish recursive feasibility of the sequential SISO loop-shaping \mc{approach, in addition to} MIMO closed-loop stability, for a model class that includes, but is not limited to, the irrigation channel model. The design \mc{method} is \mc{validated} by numerical simulations in Section \ref{sec:simulations}, \mc{and field trial results recorded across a seven-day operational end-of-season period in Section~\ref{sec:fieldtest}.} Concluding remarks are given in Section \ref{sec: conclusions}.

\section{Channel modelling}\label{sec: modelling}
\begin{figure}[tbp!]\centering
\includegraphics[width=.98\columnwidth]{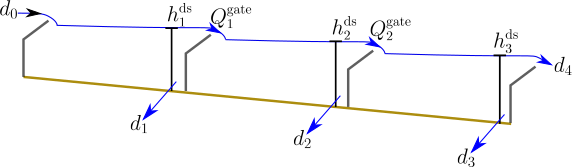}
\caption{ Channel schematic showing the downstream water levels, $\hds_n$, the flows over the controllable gates, $\Qgate_n$, and the inflow and outflow disturbances $d_n$ in a channel with $N=3$ pools.
}
\label{fig 1}
\end{figure}
A schematic of a channel  is shown in Figure \ref{fig 1}. Let $N$ denote the number of pools in which the level deviations are to be balanced. The  flows over the $N-1$ controllable gates between these pools, $\Qgate_n$,  are the control inputs.  From the perspective of  the balancing controllers, the inflow at the top of the channel, $d_0$, the outflow at the bottom of the channel, $d_{N+1}$, and the outtakes in each pool, $d_1$ through $d_N$, are treated as external disturbance inputs. The flow $d_{N+1}$, for instance, could be in practice determined by a distant downstream controller regulating the downstream pool, so it is not considered here as a control input for the purpose of level balancing. 

\subsection{Saint Venant Equations}\label{sec:Saint Venant}
\begin{figure}[tbp!]\centering
\includegraphics[width=.45\textwidth]{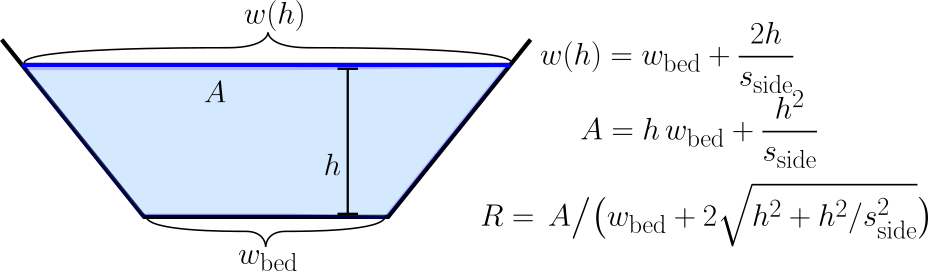}
\caption{Trapezoidal channel cross-section and relationships between wetted area $A$, water level $h$, top-width $w(h)$, bed-width $w_{\text{bed}}$, and side slope $s_{\text{side}}$.}
\label{fig:cross_section}
\end{figure}

The Saint Venant equations~\cite{chaudhry2007open,litrico2009modeling} are used to model the open-channel dynamics of each pool. \mc{The corresponding} nonlinear hyperbolic partial differential equation model \mc{is} based on the following distributed mass and momentum balances:
\begin{align}
\frac{\partial A}{\partial t} + \frac{\partial Q}{\partial x} & = 0 \label{StVenant1}\\
\frac{\partial Q}{\partial t} + \frac{\partial}{\partial x} \left(\frac{Q^2}{A} + gP \right) &= gA\left(S_0-S_f\right), \label{StVenant2}
\end{align}
where $A(x,t)$ and $Q(x,t)$ are the wetted cross-section area and flow rate, respectively, at time $t$ and location $x$. The \mc{independent} variables $x$ and $t$ may be omitted for brevity. The location $x\in[0,L]$ is measured as the distance from the upstream end, where $L$ is the length of the pool. The relations between $A$, water level $h$, channel top-width $w(h)$, and hydraulic radius $R$, are as shown in Figure \ref{fig:cross_section} for the trapezoidal channel cross-section geometry assumed throughout. The friction slope $S_f$ is modelled as
\begin{equation}
S_f(A,Q) = \frac{n_{\text{Manning}}^2Q^2}{A^2R^{4/3}},\label{eq:Sf}
\end{equation}
with Manning coefficient $n_{\text{Manning}}$. The wetted-area dependent hydrostatic pressure term\footnote{The hydrostatic pressure effect $(\partial P/\partial x)$ in
  \eqref{StVenant2} is often expressed in the form
  $(\mathrm{d}P/\mathrm{d}A) \cdot (\partial A/\partial x)$ with
  $(\mathrm{d}P/\mathrm{d}A) = A/w(h(A))$, where the dependence of $h$ on $A$ is determined by channel cross-section geometry.}
%$(\partial_{x} P)(h(A)) = (D_{\! A} P)(h(A)) \cdot (\partial_{x} A)$ where $(D_{\! A} P)(h(A))=A/w(h(A))$.}
is 
\begin{equation}
  P(A) = \int_0^{h(A)} (h(A)-\xi)w(\xi) d\xi,  \label{hydrostatic pressure}
\end{equation} 
and $g$ is gravitational acceleration. The parameters $n_{\text{Manning}}$, $w_{\text{bed}}$, $S_0$ and $s_{\text{side}}$ usually vary between pools and can in principle also vary along $x$ within each pool (although the latter is not considered here.)

The Saint Venant equations (\ref{StVenant1})-(\ref{StVenant2}) are complemented by the inflow and outflow boundary conditions
\begin{align}
Q(0,t) = \Qin(t), \qquad  Q(L,t) =  \Qout(t). \label{Saint Venant boundary condition}
\end{align}
\mc{For a channel consisting of $N$ pools, the corresponding variables $A_n$ and $Q_n$ are indexed by pool number $n\in\{1,\ldots,N\}$, and the boundary conditions become:}
\begin{align}
\Qin_n(0,t) &= \begin{cases} d_0(t),& n=1,\\ \Qgate_{n-1}(t), & n\in\{2,\ldots,N\}, \end{cases} \label{Saint Venant boundary condition 1}\\
   \Qout_n(t)& = \begin{cases} d_n(t) + \Qgate_n(t), & n\in\{1,\ldots,N-1\},\\ d_N(t) + d_{N+1}(t), & n=N,  \end{cases} \label{Saint Venant boundary condition 2}
\end{align}
\mc{where $\Qgate_{n}$ is the controlled gate flow between pool $n-1$ and $n$, and $d_{n}$ is the outtake flow in pool $n$.}

In this paper, the Saint Venant equations (\ref{StVenant1})-(\ref{StVenant2}) are used as a high-fidelity model to verify the control design in simulations in Section \ref{sec:simulations}. Moreover, the \mc{proposed sequential SISO classical loop-shaping} controller synthesis is based on \mc{distributed parameter} transfer functions \mc{that arise from} linearized Saint Venant equations \mc{for the pool dynamics}.

\begin{figure}[htbp!]\centering
\includegraphics[width=.99\columnwidth]{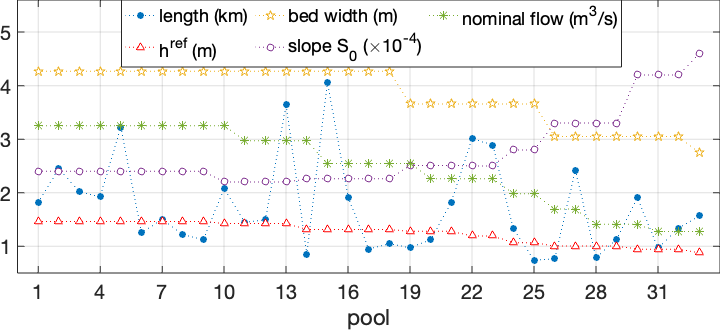}
\caption{{\red Parameters of the Saint Venant model in each pool of the exemplary channel. The quantity} $h^{\text{ref}}$ is the reference (i.e., nominal setpoint) water level at the downstream end of each pool. The  parameters  $n_{\text{Manning}}=0.0225$\,s/m$^{1/3}$ and $s_{\text{side}}=2/3$ {\red are uniform across all pools.}  }
\label{fig channel data}
\end{figure}

The \mc{parameter} data of an exemplary channel, \mc{which is} used for demonstration purposes throughout the paper, are \mc{shown} in Figure \ref{fig channel data}. The pool length is quite variable in this channel. The other parameters change more gradually, with pools getting narrower, shallower, steeper, and having a lower flow capacity further downstream in the channel.

\subsection{Frequency-domain modelling of individual pools}\label{sec frequency domain one pool}
 A frequency domain loop-shaping approach is adopted for  controller design. For this, the Saint Venant {\red model (\ref{StVenant1})-(\ref{Saint Venant boundary condition}) for each pool} is linearized around the steady state that is uniquely defined by setting the downstream water level equal to the reference value, $\hrref$, and the  flow rate to a nominal constant flow, $Q_0$. The reference  water level can be converted to a reference downstream wetted area, $\Aref$, via the relation in Figure \ref{fig:cross_section}. Then, the steady-state wetted area, $A_0$, satisfies the following ODE obtained by solving for $(\partial A/\partial x)$ with $(\partial A/\partial t)=0=(\partial Q/\partial t)$ in (\ref{StVenant1}) and (\ref{StVenant2}):
 \begin{align}
   \frac{\mathrm{d}A_0}{\mathrm{d}x}  &= g \, A_0 \, \frac{S_0-S_f(A_0,Q_0)}{g \, A_0/w(h(A_0)) - (Q_0/A_0)^2}
   % (D_{x} A_0)(x) &= g\,A_0(x)\frac{S_0-S_f(A_0(x),Q_0)}{g\, (D_{\! A} P)(h(A_0(x))) - (Q_0/A_0(x))^2},
\label{steady state Saint Venant} \\
A_0(L)&= \Aref.  
\end{align}
Define the deviation around the steady state as
\begin{align}
\tilde{A}(x,t)&=A(x,t)-A_0(x), & \tilde{Q}(x,t)&=Q(x,t)-Q_0,
\end{align}
and denote the corresponding Laplace transforms by $\hat{A}(x,s)$ and $\hat{Q}(x,s)$, respectively. The transfer functions from upstream inflow, $\hat{Q}^{\text{in}}=\hat{Q}(0,s)$, and downstream outflow, $\hat{Q}^{\text{out}}=\hat{Q}(L,s)$, to the downstream water level, $\hat{h}^{{\text{ds}}}(s)=\hat{A}(L,s)/w(\hrref)$, can be derived as shown in \cite[Sec.~3.4]{litrico2009modeling}. \mc{For reference, details of this derivation are given in the Appendix. For the purpose of subsequent developments, the model is denoted as follows:}
\begin{equation}
\hat{h}_n^{\text{ds}}(s) = G^{\text{in}}_n(s) \hat{Q}_n^{\text{in}}(s) + G^{\text{out}}_n(s) \hat{Q}_n^{\text{out}}(s), \label{transfer function pool i}
\end{equation}
where \mc{$G^{\text{in}}_n$ and $G^{\text{out}}_n$ are the corresponding distributed parameter transfer functions,} and $n\in\{1,\ldots,N\}$ is the pool index. 

\begin{figure}[tbp!]\centering
\includegraphics[width=.99\columnwidth]{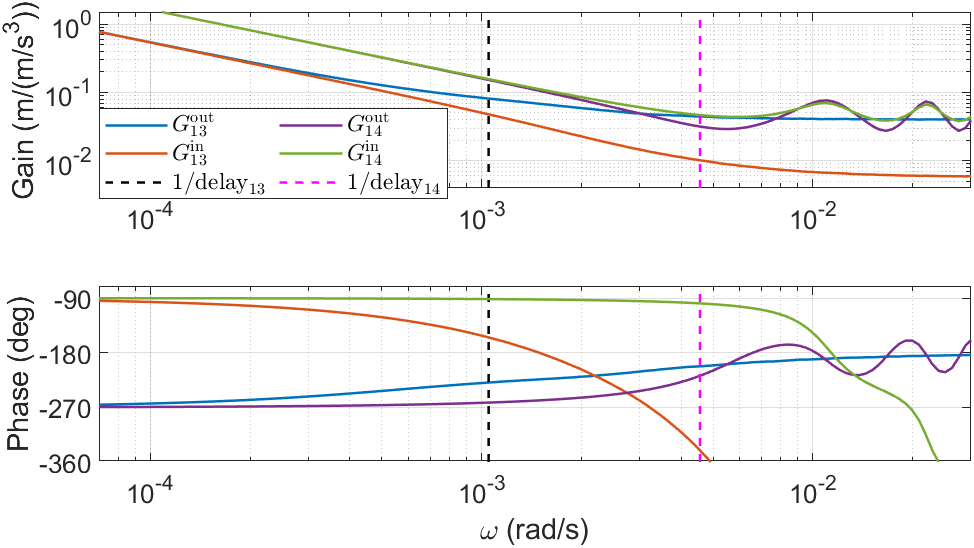}
\caption{ Bode plots of the transfer functions from in- and outflow to downstream water level for pools 13 (relatively long) and pool 14 (short).}
\label{fig Bode individual pools}
\end{figure}

The Bode plots of $G^{\text{in}}_{\mc{n}}$ and $G^{\text{out}}_{\mc{n}}$ for pools \mc{$n=13$} and \mc{$n=14$} are shown in Figure \ref{fig Bode individual pools}.  Note that
\begin{align}
\lim_{s\rightarrow 0} \left( s  \,G^{\text{in}}_n(s) \right) & = \frac{1}{c_n}, & \lim_{s\rightarrow 0} \left( s  \,G^{\text{out}}_n(s) \right) & = -\frac{1}{c_n},  \label{integrator approximation}
\end{align}
where $c_n$ {\blue (m$^2$) reflects the storage capacity of the pool per unit change of downstream water level. A larger $c_n$ indicates less sensitivity of the water level to mismatch between in- and outflow}. In particular, for constant in- and outflows, the levels in the pool rise or fall at a rate {\blue that is proportional to the reciprocal of the} pool capacity. For the {\blue example channel, the} integrator gain {\blue $1/c_n$ in pool} $n=13$ is smaller than \mc{in pool} $n=14$, \mc{since the former is of the same width but longer}. There is a  delay before inflows over the upstream gate reach the downstream end of the pool, which dominates the phase of $G^{\text{in}}_{\mc{n}}$ at higher frequencies starting at approximately the inverse of that delay. \mc{On the other hand,} $G^{\text{out}}_{\mc{n}}$ is not affected by any delay and \mc{it} has a zero that increases the phase towards $-180$\,deg \mc{at high frequency}. Moreover, wave effects are clearly visible for pool 14, whereas in pool 13 (longer) friction is more dominant, which dampens the resonances, as reflected in the smoother Bode magnitude plot at high frequencies.

\begin{remark}\label{remark:poles}
In general, it is only possible to compute the poles numerically \cite[Sec.~3.4]{litrico2009modeling}. In very special cases, the poles can be found analytically~\cite[Sec.~3.1.1]{litrico2009boundary}, \mc{but rarely in practice. On the other hand,} it can be shown that if friction is non-negligible, then \mc{qualitatively}: \mc{1) } all poles have negative real part, {\red except for the \mc{simple} pole at $s=0$}\mc{;} \mc{2)} there is a finite number of negative real poles\mc{;} and \mc{3)} all other \mc{poles arise in} complex conjugate pairs that represent decaying oscillations.
\end{remark}

\subsection{Interconnected model structure}

\begin{figure}[tbp!]\centering
\includegraphics[width=.95\columnwidth]{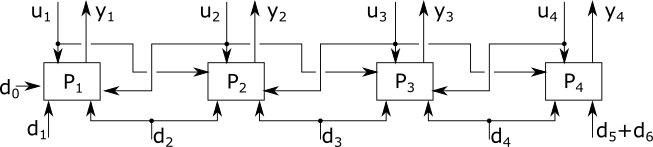}
\caption{Structure of the uncontrolled system (\ref{y MIMO}). The figure shows an example with $N=5$ pools and $N-1=4$ subsystems and control inputs.}
\label{fig 3}
\end{figure}

The input to \mc{the} local controller \mc{at gate $n\in\{1,\ldots,N-1\}$} is the difference between the water-level errors in the two pools upstream and downstream of the gate: 
\begin{align}
y_n(t) &= \left(\hrref_{n} - {h}_{n}^{\ds}(t)\right) - \left(\hrref_{n+1}- {h}_{n+1}^{\ds}(t)\right). \label{y definition} % (t), not (s)
\end{align} 
The controller output is the corresponding gate flow  
\begin{align}
u_n(t)&=\Qgate_{n}(t),& n\in\{1,\ldots,N-1\}.  \label{u definition}
\end{align}
\mc{In practice, a faster timescale lower-level controller would be employed to adjust the gate position, relative to  measured water level upstream and/or downstream of the gate, depending on type, to deliver commanded gate flow $u_n$.} The supply inflow $d_0$, off-takes $d_n$, $n\in\{1,\ldots,N\}$, and downstream outflow $d_{N+1}$ are external disturbances. See Figure \ref{fig 1}. Let $\bm{u}=(u_1,\ldots,u_{N-1})$, $\bm{y}=(y_1,\ldots,y_{N-1})$ and $\bm{d}=(d_0,\ldots,d_{N+1})$; more generally, the notation $\bm{a}=(a_1,\ldots,a_n)$ denotes a column vector $\bm{a}$ with ordered scalar entries $a_1,a_2,\ldots,a_n$, where $n=\mathrm{dim}(\bm{a})$ is the dimension. Using (\ref{Saint Venant boundary condition})-(\ref{Saint Venant boundary condition 2}) and (\ref{transfer function pool i}) in (\ref{y definition}), and denoting the Laplace transforms of $\bm{u}$, $\bm{y}$ and $\bm{d}$ by $\hat{\bm{u}}$, $\hat{\bm{y}}$ and $\hat{\bm{d}}$, respectively, the  open-loop system  can be written in  MIMO form as
\begin{equation}
\hat{\bm{y}} = \bm{G} \hat{\bm{u}} + \bm{G}^d \hat{\bm{d}}  \label{y MIMO}
\end{equation}
with
\begin{align}
G_{n,n} &=G^{\text{in}}_{n+1} - G^{\text{out}}_{n}, & n&\in\{1,\ldots,N-1\}, \\
G_{n,n+1} &= G^{\text{out}}_{n+1}, &  n&\in\{1,\ldots,N-2\}, \\
G_{n,n-1} &= -G^{\text{in}}_{n}, &  n&\in\{2,\ldots,N-1\}, \\
%\intertext{and}
G^d_{1,1} &= -G^{\text{in}}_{1},\\
G^d_{n,n+1} &= -G^{\text{out}}_{n}, & n&\in\{1,\ldots,N-1\},\\
G^d_{n,n+2} &= G^{\text{out}}_{n+1}, & n&\in\{1,\ldots,N-1\},\\
G^d_{N-1,N+2} &= G^{\text{out}}_{N},
\end{align}
where $G_{k,l}$ denotes the SISO transfer function in row $k$, column $l$ of $\bm{G}$, and all other entries equal to zero.
That is, $\bm{G}$ is a tri-diagonal $(N-1) \times (N-1)$ transfer matrix, and $\bm{G}^d$ is a similarly sparse transfer matrix. As such, a channel consisting of $N$ pools with the control inputs (\ref{u definition}) and  measured outputs  (\ref{y definition})  can be seen as a string of $N-1$ subsystems that are interconnected according to  the structure  depicted in Figure \ref{fig 3}.  There, $P_1$ is the local system with output $y_1$, which depends on inputs $u_1$ and $u_2$,  disturbances $d_0$, $d_1$ and $d_2$, and the dynamics in both pools 1 and 2. System $P_2$ with output $y_2$ depends on pools 2 and 3, etc. 

\begin{figure}[tbp!]\centering
\includegraphics[width=.99\columnwidth]{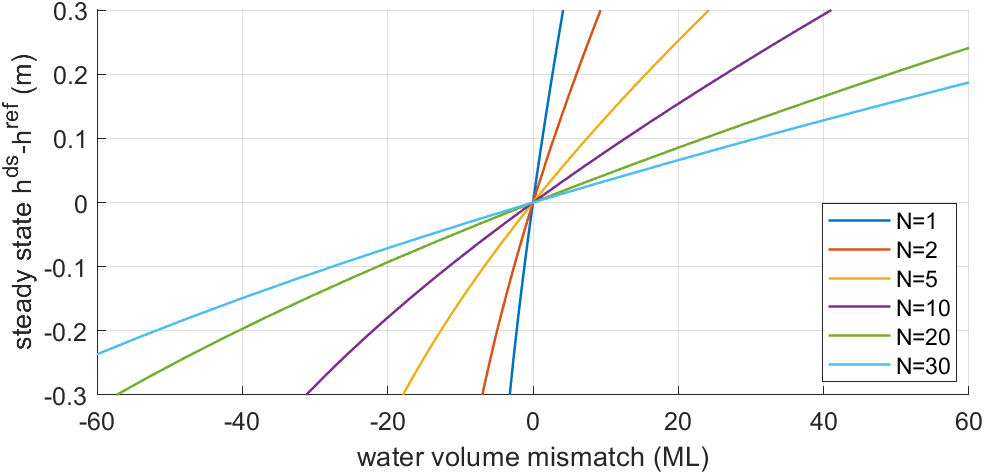}
\caption{Water level error (m) in each pool \mc{versus} the  water volume (ML) mismatch,  assuming that all pools are balanced and at equilibrium with $y_n=0$ for all $n\in\{1,\ldots,N-1\}$ and $d_0=\sum_{n=1}^{N+1}d_n$, for the first $N$ pools of the channel from Figure \ref{fig channel data}. The point $(0,0)$ corresponds to all pools at reference downstream water level. }
\label{fig dV to dh }
\end{figure}

If a controller is able to regulate the outputs $y_n$ to zero in steady-state, i.e.,  $y_n=0$ for all $n\in\{1,\ldots,N-1\}$, then $y_1=0$ and (\ref{y definition}) directly implies $\hds_1-\hrref_1=\hds_2-\hrref_2$. Similarly, $y_1=y_2=0$ implies $y_1+y_2 =  (\hrref_{1} - h_{1}^{\ds}) - (\hrref_{2}- h_{2}^{\ds}) +  (\hrref_{2} - h_{2}^{\ds}) - (\hrref_{3}- h_{3}^{\ds}) =0$, i.e.,  $\hds_1-\hrref_1=\hds_3-\hrref_3$. Repeating this argument leads to 
 \begin{align}
&\left(y_n=0 \text{ for all }  n\in\{1,\ldots,N-1\}  \right)  \Leftrightarrow \nonumber \\ 
&\left( \hds_n-\hrref_n=\hds_m-\hrref_m  \text{ for all }  n,m\in\{1,\ldots,N-1\} \right).
 \end{align}
 Since there is one less output than pools, $y_n=0$ for all $n\in\{1,\ldots,N-1\}$, is not sufficient to define an equilibrium for the water levels. For instance, the water-level errors in all pools could be equal and  rise at the same rate if the top inflow exceeds the sum of all off-takes. However, the water levels can reach an equilibrium  if additionally the inflow  matches the off-takes or, in other words, if the total water volume in all balanced pools is constant. In this case, the total water volume \mc{can} be computed via \mc{spatial integration of the solution $A_0$ of the} steady-state model (\ref{steady state Saint Venant}) with appropriate downstream boundary condition \mc{and compared to} the water volume with all pools at their reference levels. The resulting relationship between the water volume mismatch and the level tracking error in each pool is shown in Figure \ref{fig dV to dh }. Clearly, a larger number of balanced pools significantly reduces the sensitivity to mismatch. The slightly nonlinear curvature is due to the trapezoidal cross section and nonlinear water surface profile. Similarly, in the case that $d_0\neq\sum_{n=1}^{N+1}d_n$, the average water levels rise or fall more slowly with a larger  combined storage capacity in the balancing pools. Also see Remark \ref{remark integrator gain d to h} later in this paper.
 %, and the comparison between Figures \ref{fig simulation decentralized 20} and \ref{fig simulation decentralized 10}.

\section{Decentralized control design} \label{sec: control design}
\begin{figure}[htbp!]\centering
\includegraphics[width=.95\columnwidth]{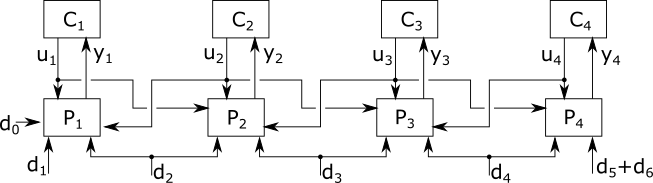}
\caption{Structure of the closed-loop system with decentralized feedback control (omitting feedforward paths for clarity).}
\label{fig 2}
\end{figure}
In decentralized feedback control, each control input $u_n$ is determined based on only the local measurement $y_n$.  This  feedback controller structure is shown in Figure \ref{fig 2}. However, due to the bi-directional coupling between the subsystems, each $u_n$ also affects all other outputs $y_m$,  $m\neq n$. In other words, due to the tri-diagonal structure of $\bm{G}$ in  (\ref{y MIMO}),  closing the loop  with $\hat{\bm{u}}=\bm{C}\hat{\bm{y}}$ and $\bm{C}=\operatorname{diag}(C_1,\ldots,C_{N-1})$ 
diagonal results in a full (i.e., unstructured) inverse $(\bm{I}-\bm{G}\bm{C})^{-1}$. As such, naively designing each SISO controller $C_n$ to stabilize only the local transfer function $G_{n,n}$ from $u_n$ to $y_n$ without accounting for the effect of the other controllers via the coupling terms, would not lead to stability of the overall (MIMO) system in general.  \mc{With this in mind, the} sequential design procedure developed below \mc{takes into} account the control loops that have been closed in \mc{preceding} steps to ensure MIMO stability, whilst maintaining the advantage that each step involves only SISO classical loop-shaping design.

\subsection{Definition of the sequential SISO design procedure}\label{subsection recursion}

\begin{figure}[htbp!]\centering
\includegraphics[width=.95\columnwidth]{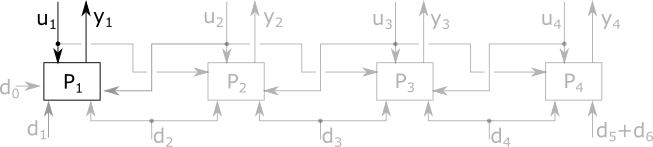}\\
\includegraphics[width=.95\columnwidth]{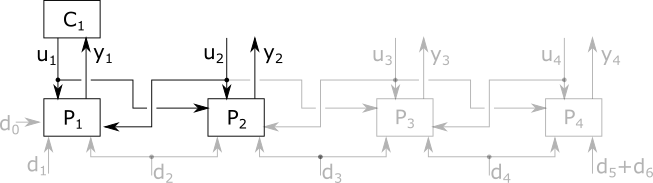}\\
\includegraphics[width=.95\columnwidth]{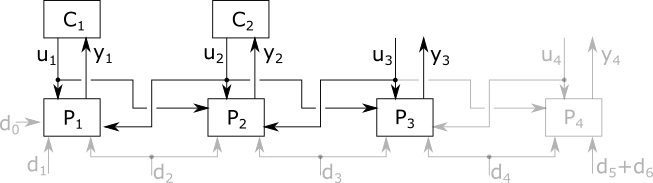}
\caption{Model structure during the first three recursive loop-shaping steps if the controllers are designed in the order $\bm{\nu} = (1,2,3,\ldots)$. Parts that have no effect on the dynamics from $u_{\nu_m}$ to  $y_{\nu_m}$ at step $m$ are shaded grey. }
\label{fig recursive loop-shaping}
\end{figure}
\begin{figure}[htbp!]
\includegraphics[width=.95\columnwidth]{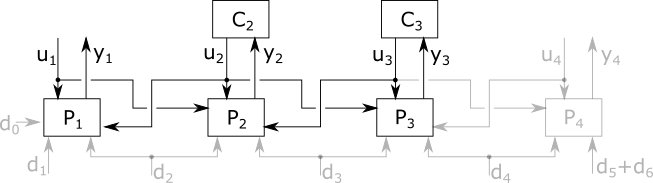}\hfill \vspace{.2cm}\\
\includegraphics[width=.95\columnwidth]{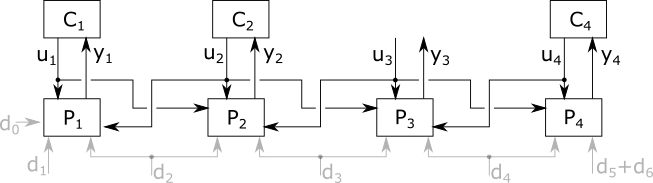}\hfill 
\caption{Top: Model structure from control input $u_1$ to measurement $y_1$ during the third recursive loop-shaping step for  $\bm{\nu} = (2,3,1,4)$, i.e., after controllers $C_2$ and $C_3$ are already designed. Bottom: Model structure during the fourth step for $\bm{\nu} = (1,4,2,3)$.}
\label{fig recursive loop-shaping 2}
\end{figure}

Let $\bm{\nu}=(\nu_1,\nu_2,\ldots,\nu_{N-1})$ be any permutation of $\{1,\ldots,N-1\}$ corresponding to the gate indices in the order that the decentralized controllers are to be designed. Graphically, the plant `seen' by the controller to be designed at each recursive step is illustrated in Figures \ref{fig recursive loop-shaping} and \ref{fig recursive loop-shaping 2} for different choices of the order. At step $m\in\{1,\ldots,N-1\}$, the controller $C_{\nu_m}$ for gate $\nu_m$ is designed for the transfer function from  $\hat{u}_{\nu_m}$ to $\hat{y}_{\nu_m}$ with the control loops $\hat{u}_{\nu_p}=C_{\nu_p}\hat{y}_{\nu_p}$ of all previous steps $p\in\{1,\ldots,m-1\}$ in place (i.e.,  already closed).

The transfer function from  $\hat{u}_{\nu_m}$ to $\hat{y}_{\nu_m}$ of the partially closed-loop system at design step $m\in\{1,\ldots,N-1\}$ is derived next. To this end, the following notation is used: 
\begin{itemize}
\item Let $\underline{\bm{\nu}}^{[m]}=(\nu_1,,\ldots,\nu_{m-1})$ and $\overline{\bm{\nu}}^{[m]}=(\nu_{m+1},\ldots,\nu_{N-1})$ denote the indices of control loops that have been closed in antecedent steps, and those that will be closed in subsequent steps, respectively. 
\item Given $k,l\in\{1,\ldots, N-1\}$, and index vectors $\bm{\gamma}=(\gamma_1,\ldots, \gamma_k)\subset\{1,\ldots,N-1\}$ and $\bm{\delta}=(\delta_1,\ldots, \delta_l)\subset\{1,\ldots,N-1\}$, with $\gamma_i \neq \gamma_j$ and
$\delta_i\neq \delta_j$ if $i\neq j$, the $k\times l$ transfer function matrix $\bm{G}_{\bm{\gamma},\bm{\delta}}=\bm{H}$ with
$H_{i,j}=G_{\gamma_i,\delta_j}$ for $i\in\{1,\ldots,k\}$ and $j\in\{1,\ldots,l\}$, consists of the entries of $\bm{G}$ specified by $\bm{\gamma}$ and $\bm{\delta}$; 
\item The $k$-th row, and respectively $k$-th column of $\bm{G}$, are denoted by $\bm{G}_{k,\bullet}$ and $\bm{G}_{\bullet,k}$ for $k\in\{1,\ldots,N-1\}$;
\item Given $\bm{a}=(a_1,\ldots,a_{N-1})$ and index vector $\bm{\gamma}=(\gamma_1,\ldots, \gamma_k)\subset\{1,\ldots,N-1\}$, the sub-vector permutation $\bm{a}_{\bm{\gamma}}$ denotes $(a_{\gamma_1},a_{\gamma_2},\ldots,a_{\gamma_k})$;
\item 
$\tilde{\bm{G}}^{[m]}$ is defined by
\begin{equation}
\tilde{G}^{[m]}_{k,l} = \begin{cases} G_{k,l} & \text{if } l\notin \underline{\bm{\nu}}^{[m]} \\ 0 & \text{if } l\in \underline{\bm{\nu}}^{[m]} \end{cases}, \quad k,l\in\{1,\ldots,N-1\}.  \label{Gtilde}
\end{equation}
\end{itemize}
Given this notation, the  open-loop system (\ref{y MIMO}) 
can be rewritten as
\begin{align}
\left(\begin{matrix}
 \hat{\bm{y}} \\  \hat{\bm{y}}_{\underline{\bm{\nu}}^{[m]}}
\end{matrix} \right) = & 
\left(\begin{matrix}
\tilde{\bm{G}}^{[m]} &  \bm{G}^d & \bm{G}_{\bullet,\underline{\bm{\nu}}^{[m]}}\\
\tilde{\bm{G}}^{[m]}_{\underline{\bm{\nu}}^{[m]}\!,\bullet}  & \bm{G}^d_{\underline{\bm{\nu}}^{[m]}\!,\bullet} & \bm{G}_{\underline{\bm{\nu}}^{[m]}\!,\underline{\bm{\nu}}^{[m]}}
\end{matrix} \right)
\left(\begin{matrix}
 \hat{\bm{u}} \\  \hat{\bm{d}} \\ \hat{\bm{u}}_{\underline{\bm{\nu}}^{[m]}}
\end{matrix}\right).
\end{align}
The controllers already in place from previous steps correspond to 
\begin{equation}
\hat{\bm{u}}_{\underline{\bm{\nu}}^{[m]}} = {\bm{C}}_{\underline{\bm{\nu}}^{[m]}} \hat{\bm{y}}_{\underline{\bm{\nu}}^{[m]}} \label{some loops closed}
\end{equation}
with
\begin{equation}
{\bm{C}}_{\underline{\bm{\nu}}^{[m]}} = \operatorname{diag}
\left(C_{\nu_1},C_{\nu_2},\ldots,C_{\nu_{m-1}}\right). \label{C underline step i}
\end{equation}

At  step $m\in\{1,\ldots,N-1\}$, the transfer function from the target control input $\hat{u}_{\nu_m}$ of the current step, the control inputs $\hat{\bm{u}}_{\bar{\bm{\nu}}^{[m]}}$ to be designed in subsequent steps, and the external disturbances  $\hat{\bm{d}}$, to the  \mc{measurement} outputs \mc{permuted} according to \mc{$({\nu_m},  {\overline{\bm{\nu}}^{[m]}},  {\underline{\bm{\nu}}^{[m]}})$}, is given by
\begin{equation}
\left(\begin{matrix}
 \hat{y}_{\nu_m} \\ \hat{\bm{y}}_{\overline{\bm{\nu}}^{[m]}} \\ \hat{\bm{y}}_{\underline{\bm{\nu}}^{[m]}}
\end{matrix} \right)
= 
\left(\begin{matrix} 
G^{[m]} & \bm{H}^{[m]}_{\nu_m,\smash{\overline{\bm{\nu}}^{[m]}}} & \bm{J}^{[m]}_{\nu_m,\bullet} \\
\bm{H}^{[m]}_{\overline{\bm{\nu}}^{[m]}\!,\nu_m} & \bm{H}^{[m]}_{\overline{\bm{\nu}}^{[m]},\overline{\bm{\nu}}^{[m]}} & \bm{J}^{[m]}_{\overline{\bm{\nu}}^{[m]}\!,\bullet} \\
\bm{H}^{[m]}_{\underline{\bm{\nu}}^{[m]}\!,\nu_m} & \bm{H}^{[m]}_{\underline{\bm{\nu}}^{[m]}\!,\overline{\bm{\nu}}^{[m]}} & \bm{J}^{[m]}_{\underline{\bm{\nu}}^{[m]}\!,\bullet} 
 \end{matrix}  \right) 
 \left(\begin{matrix} \hat{u}_{\nu_m} \\  \hat{\bm{u}}_{\overline{\bm{\nu}}^{[m]}} \\ \hat{\bm{d}} \end{matrix}   \right) \label{G partially closed loop}
\end{equation}
with
\begin{align}
&\bm{H}^{[m]} \nonumber \\
&= \tilde{\bm{G}}^{[m]} + \bm{G}_{\bullet,\underline{\bm{\nu}}^{[m]}}\bm{C}_{\underline{\bm{\nu}}^{[m]}}\left(\bm{I}-\bm{G}_{\underline{\bm{\nu}}^{[m]}\!,\underline{\bm{\nu}}^{[m]}}\bm{C}_{\underline{\bm{\nu}}^{[m]}} \right)^{-1}\tilde{\bm{G}}^{[m]}_{\underline{\bm{\nu}}^{[m]}\!,\bullet},
\label{H definition} \\
&\bm{J}^{[m]} \!=\! \bm{G}^d + \bm{G}_{\bullet,\underline{\bm{\nu}}^{[m]}}\bm{C}_{\underline{\bm{\nu}}^{[m]}}\left(\bm{I}-\bm{G}_{\underline{\bm{\nu}}^{[m]}\!,\underline{\bm{\nu}}^{[m]}} \bm{C}_{\underline{\bm{\nu}}^{[m]}}\right)^{-1}\bm{G}_{\underline{\bm{\nu}}^{[m]}\!,\bullet}^d, \label{J definition} \\
&G^{[m]} \!=\! \bm{H}^{[m]}_{\nu_m,\nu_m}. \label{G^[i] definition}
\end{align}
The controller synthesis problem entails the design of the SISO compensator $C_{\nu_m}$ to achieve a desirable local loop transfer function $G^{[m]} C_{\nu_m} $.

\subsection{Open-loop transfer functions and controller design}\label{sec: open loop transfer functions}
\begin{figure}[htbp!]\centering 
\includegraphics[width=.99\columnwidth]{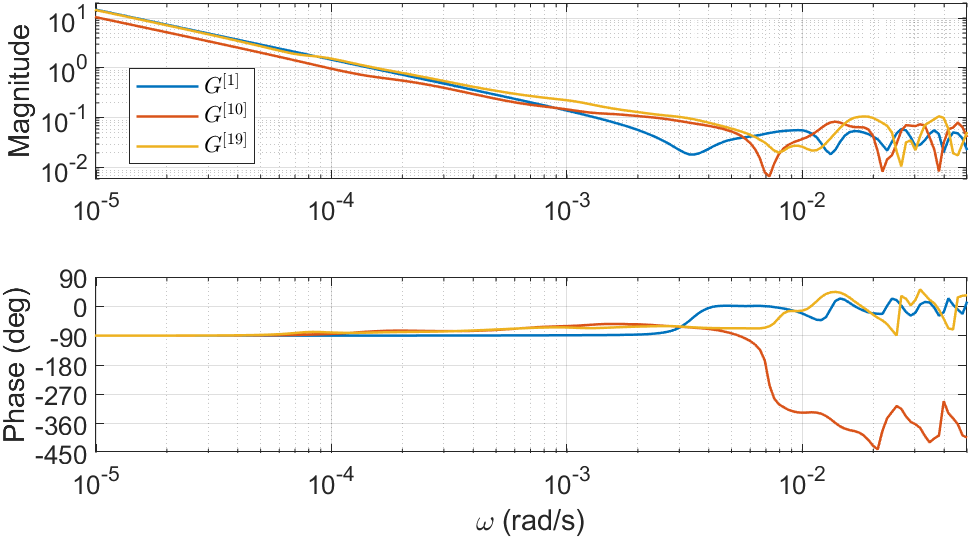}
\caption{The open-loop transfer functions $G^{[m]}$ from $\hat{u}_{\nu_m}$ to $\hat{y}_{\nu_m}$ for $\bm{\nu}=(1,2,\ldots,19)$ and $m=1,10,19$.}
\label{fig open loop G}
\end{figure}

 Figure \ref{fig open loop G} shows the transfer function $G^{[m]}$ for steps $m\in\{1,10,19\}$ in the case where controllers for the first 20 pools of a channel, with the characteristics shown in Figure \ref{fig channel data}, are designed in the order $\bm{\nu}=(1,2,\ldots,19)$. As for the individual pool transfer functions in Section \ref{sec frequency domain one pool}, the integrator dynamics ($-90$\,deg phase, $20$dB/decade roll-off) dominate at low frequencies. A rigorous derivation of the gain $\lim_{s\rightarrow 0} |s\,G^{[m]}|$ is presented in Section~\ref{sec proofs}. At higher frequencies, the integrator-like magnitude roll-off stops, and resonance/anti-resonances can be seen. The phase plot is less structured and depends on the pool dimensions relative to each other and on the order $\bm{\nu}$,  among other things. In many cases, like ${G}^{[1]}$ and ${G}^{[19]}$ in Figure~\ref{fig open loop G}, the phase increases to above $-90$\,deg in the frequency range where the magnitude roll-off stops, resembling the effect of a left-half plane zero, making the corresponding dynamics passive. In other cases, like $G^{[10]}$ shown in Figure~\ref{fig open loop G}, there is non-passive delay-like dynamics at play, requiring more care when closing the loop. 
 
 PI compensators of the form
\begin{equation}
C_n(s) = -K^p_n\left(1 + \frac{1}{T^I_n\,s} \right) \frac{1}{1+T^F_n\,s}, \label{C_i form}
\end{equation}
are used for loop-shaping, with $K^p_n>0$, $T^I_n>0$, and low-pass filter time constant $T^F_n>0$. The integrator term serves to achieve zero tracking error differences at steady state in   presence of constant off-take flow disturbances, as established in Theorem \ref{theorem MIMO stability} further below.  The low-pass filter ensures that the controller rolls off at higher frequencies, in order to i) reduce sensitivity to uncertainty in the high-frequency dynamics, ii) avoid exciting wave resonances, and iii) reduce actuator wear. The design of such compensators for closed-loop control is well studied, e.g., see~\cite{astrom2006pid,astrom2021feedback}. 

\begin{figure}[htbp!]\centering 
\includegraphics[width=.99\columnwidth]{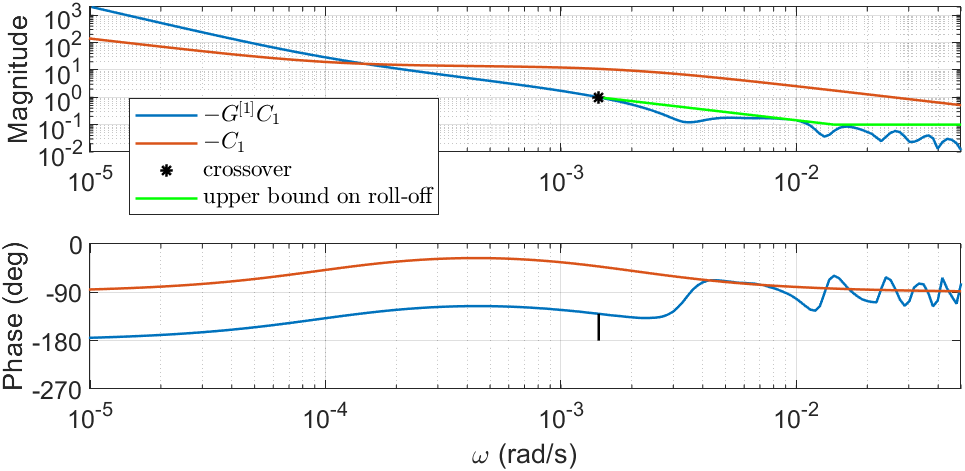}\\
\includegraphics[width=.99\columnwidth]{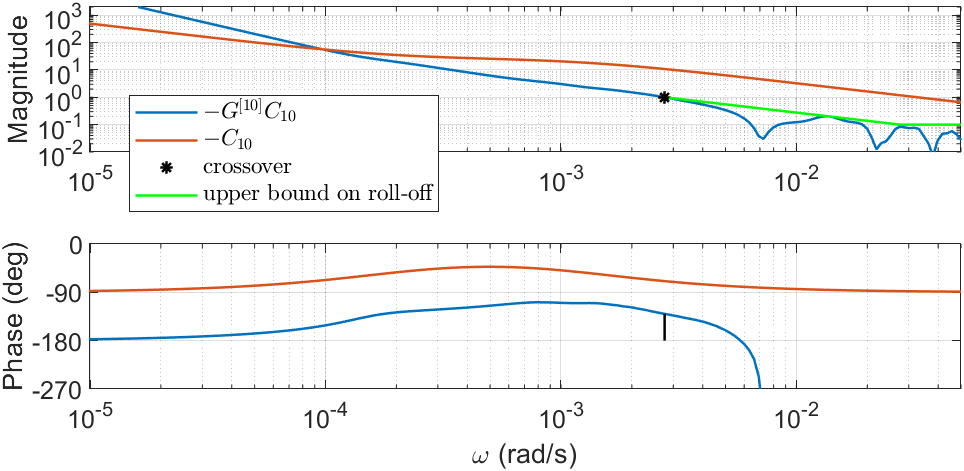}\\
\includegraphics[width=.99\columnwidth]{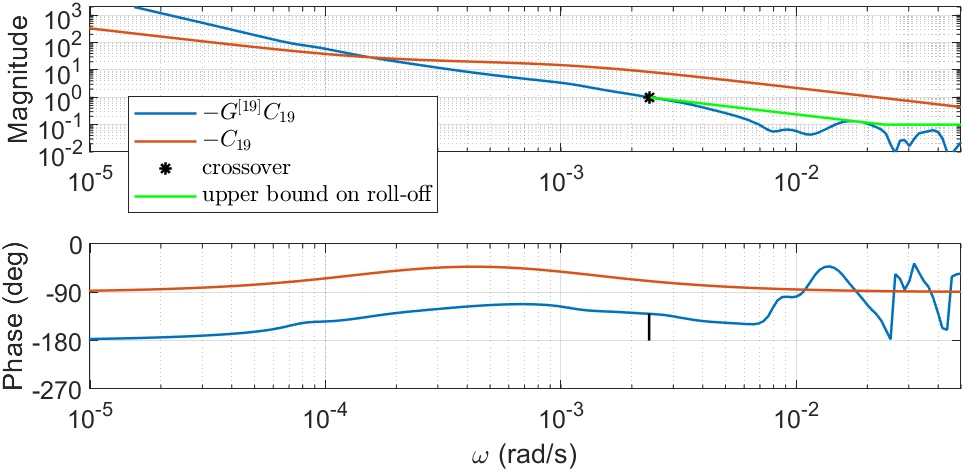}
\caption{ Transfer functions of the controllers $C_{\nu_{m}}$ (in units m$^3$/s/m) and the shaped loops ${G}^{[m]}C_{\nu_{m}}$ during the first, 10-th and 19-th recursive loop-shaping steps, for $\bm{\nu}=(1,2,\ldots,N-1)$, $N=20$.  }
\label{fig shaped GK}
\end{figure}

Combined, the integrator-like dynamics of $G^{[m]}$ with the integrator in $C_{\nu_m}$, means the \mc{phase of the} shaped transfer function $-G^{[m]} C_{\nu_m}$ %has 
\mc{is} close to $-180$\,deg phase at low frequencies. In order to achieve stability and a good phase margin, the phase of $-G^{[m]} C_{\nu_m}$ at the unit magnitude crossover frequency needs to be sufficiently greater than $-180$\,deg. If the phase of $G^{[m]}$ falls significantly below $-90$\,deg at high frequencies, then the achievable crossover frequency is limited. In other cases, crossover should be placed somewhat below the resonances and other uncertain high-frequency dynamics. The phase lag of the PI controller at crossover can be reduced by placing the zero of the PI controller sufficiently to the left of crossover. While use of the low-pass filter is not strictly necessary, it can improve roll-off after crossover. Without this filter, a larger margin between crossover and the first resonant frequency might be required to ensure a satisfactory gain margin. A higher crossover frequency is desirable for fast disturbance rejection. The phase lag of the low-pass filter can be handled by leaving enough separation between crossover and the pole of the low-pass filter, although this might not be required if the phase of $G^{[m]}$ at crossover is well above -$90$\,deg. In summary, the SISO controller at each step is shaped based on the following criteria:
\begin{enumerate}
\item The local SISO \mc{(positive feedback)} loop consisting of $G^{[m]}$ and $C_{\nu_m}$ is closed-loop stable with a specified phase margin $\phi_{\text{PM}}\in(0,90)$\,deg \mc{(i.e., $\angle -G^{[m]}C_{\nu_m}=-180+\phi_{\text{PM}}\in(-180,-90)$\,deg at the crossover frequency where $|G^{[m]}\,C_{\nu_m}|=1$)};
\item Beyond the crossover frequency, the magnitude $|G^{[m]}\,C_{\nu_m}|$ continues rolling off until it falls below -$20$dB, and then remains below -$20$dB at higher frequencies;
\item Subject to 1) and 2), and the preference to have crossover lie closer to the pole of the low-pass filter than to the zero of the PI controller, the crossover frequency should be  `maximized', but not beyond the inverse of the delay to avoid exciting resonances. 
\end{enumerate}
Figure \ref{fig shaped GK} shows the shaped controllers and loop transfer functions for the same PI compensator synthesis order $\bm{\nu}=(1,2,\dots, 19)$ as the sequence of open-loop transfer functions shown in Figure \ref{fig open loop G}. At step $m=10$, the phase drop of $\bar{G}^{[10]}$ puts a limit on the maximum achievable crossover frequency. For the other cases, the phase of $G^{[m]}$ is less of a limitation. However, the resonances require $T^F$ to be sufficiently small in order to satisfy the roll-off condition, and crossover is then placed at a frequency where the resonances are not excited and the phase condition is satisfied.

\subsection{Proof of recursive feasibility and MIMO stability}\label{sec proofs}
The proposed sequential SISO loop-shaping procedure is analyzed in this section. More specifically, well-posedness of this recursive loop-shaping procedure is established in the sense that $G^{[m]}$ in~\eqref{G^[i] definition}, at each step $m\in\{1,\ldots, N-1\}$, has a structure for which a stabilizing local controller $C_{\nu_m}$ of the form (\ref{C_i form}) is guaranteed to exist. Moreover, it is shown that the {\red sequentially} designed controllers $C_{\nu_m}$  collectively achieve overall MIMO closed-loop system stability, and that the controllers asymptotically reject the effect of constant disturbances.

\begin{lemma}\label{lemma integrator gain}
Given decentralized compensator design order $\bm{\nu}=(\nu_1,\ldots,\nu_{N-1})$, let 
\begin{align}
\mu^{\us}_{m}&=
\mc{\nu_m - \max} \{~n   ~:~  
\mc{{\textstyle \bigcup_{0\leq k\leq n}} \{ \nu_m - k \} }
\nonumber \\
&\qquad\qquad\qquad\qquad\qquad\subseteq 
\{\nu_1,\nu_2,\ldots,\nu_m\}  ~ \}, \nonumber\\
\mu^{\ds}_{m}&=
\mc{\nu_m} + \max\{~n ~:~ 
\mc{{\textstyle \bigcup_{0\leq k\leq n}} \{ \nu_m + k \} }
\nonumber \\
&\qquad\qquad\qquad\qquad\qquad\subseteq \{\nu_1,\nu_2,\ldots,\nu_m\} ~\},
\end{align}
for $m\in\{1,\ldots,N-1\}$. Then, then \mc{$G^{[m]}$ has a simple pole at $s=0$, and the corresponding} integrator gain is
\begin{equation}
\lim_{s\rightarrow 0} \left(s\,G^{[m]} \right) = 
1\Big/(\sum_{n=\mu^{\us}_{m}}^{\nu_m} c_n)
~+~ 
1\Big/(\sum_{n=\nu_m+1}^{\mu^{\ds}_{m}\mc{+1}} c_n)
\label{integrator step i}
\end{equation}
with each $c_n$ as per~\eqref{integrator approximation}.
\end{lemma}
\begin{remark}
In Lemma \ref{lemma integrator gain}, $\mu^{\us}_{m}$ is the \mc{index} of the \mc{gate} most upstream \mc{from} gate $\nu_m$ \mc{with a} controller in place \mc{at step $m$}, along with controllers for all gates in between, \mc{unless $\mu^{\us}_m=\nu_m$, when there is no such gate.} Similarly, $\mu^{\ds}_{m}$ \mc{is the index of} the gate most downstream \mc{from} $\nu_m$ \mc{with} a \mc{controller in place at step $m$, along with controllers for all gates in between}, \mc{unless $\mu^{\ds}_{m}=\nu_m$, when there is no such gate.} 
\end{remark}
\begin{proof}
\mc{From the definition of $\hat{y}_{\nu_{m}}$} in (\ref{y definition}), $G^{[m]}$ in (\ref{G^[i] definition}) is equal to the (partially closed-loop) transfer function from $\hat{u}_{\nu_{m}}$ to $\hat{h}^{\text{ds}}_{\nu_{m}+1}$, minus the transfer function from $\hat{u}_{\nu_{m}}$ to $\hat{h}^{\text{ds}}_{\nu_{m}}$. \mc{These transfer functions each have an isolated pole at the origin.} For brevity, \mc{this fact} and the integrator gain from $\hat{u}_{\nu_{m}}$ to $\hat{h}^{\text{ds}}_{\nu_{m}+1}$ are derived \mc{to yield} the second term in (\ref{integrator step i}); \mc{to obtain the first term, analogous} steps can be repeated for the transfer function to $-\hat{h}^{\text{ds}}_{\nu_{m}}$. 

\mc{First recall by Remark~\ref{remark:poles} and (\ref{integrator approximation}) that for $n\in\{1,\ldots,N\}$,
\begin{align*}
  G^{\text{in}}_{n}(s) = \frac{1}{s\,c_{n}} \, F^{\text{in}}_{n}(s), ~\text{ and }~ G^{\text{out}}_{n}(s) = \frac{-1}{s\,c_{n}}\,  F^{\text{out}}_{n}(s), 
\end{align*}
where $F^{\text{in}}_{n}(0)=
F^{\text{out}}_{n}(0)=1$. Further, at design step $m\in\{1,\ldots, N-1\}$, let $\hat{\bm{d}}=0$ and $\hat{\bm{u}}_{\overline{\bm{\nu}}^{[m]}}=0$ in~\eqref{G partially closed loop}, to align with the focus on $G^{[m]}$ here.  

If $\mu_{m}^{\text{ds}}=\nu_m$, then  
$G^{[m]} = G^{\text{in}}_{\nu_m+1}(s)$, in line with the corresponding second term $1/c_{\nu_m+1}$ in~\eqref{integrator step i} for this case. Otherwise, one or more consecutive controllers are in place downstream of $\nu_m$. Starting with $n=\mu_{m}^{\text{ds}} > \nu_m$, 
\begin{align} \label{eq:closingtheloopA}
    \hat{u}_{n}(s) = C_{n}(s)\,
    \big(\hat{h}_{n+1}^{\ds}(s) -  \hat{h}_{n}^{\ds}(s)\big)
\end{align}
where 
\begin{align} \label{eq:closingtheloopB}
\begin{split}
&\begin{bmatrix}
\hat{h}_{n}^{\ds}(s)\\
\hat{h}_{n+1}^{\ds}(s)
\end{bmatrix} \\
&=
\begin{bmatrix}
\frac{1}{s\,c_{n}} & 0  \\
0 & \frac{1}{s\,\gamma_{n+1}}  \\
\end{bmatrix}
\begin{bmatrix}
-F^{\text{out}}_{n}(s) & F^{\text{in}}_{n}(s) \\
\Phi^{\text{in}}_{n+1}(s) & 0 
\end{bmatrix}
\begin{bmatrix}
\hat{u}_{n}(s)\\
\hat{u}_{n-1}(s)
\end{bmatrix},
\end{split}
\end{align}
with in this case,
$\Phi_{n+1}^{\text{in}} = F_{n+1}^{\text{in}}$ and  $\gamma_{n+1}=c_{n+1} = c_{\mu_{m}^{\ds}+1}$;
n.b., $\hat{u}_{n+1}=0$, and $\Phi_{n+1}^{\text{in}}(0)=1$ 
for $n=\mu_{m}^{\ds}$. It follows that 
\begin{align*}
&\hat{u}_{n}(s)=- E_{n}(s) C_{n}(s) \frac{1}{s\,c_{n}}  F^{\text{in}}_{n}(s) \hat{u}_{n-1}(s)
\end{align*}
where 
\begin{align*}
    E_{n}(s) = \left(1 - C_{n}(s)\Big(\frac{1}{s\,c_{n}}  F^{\text{out}}_{n}(s)  + \frac{1}{s\, \gamma_{n+1}} \Phi^{\text{in}}_{n+1}(s) \Big)\right)^{-1},
\end{align*}
and thus,
\begin{align*}
   \hat{h}_{n}^{\ds}(s) =\frac{1}{s \, c_{n}}
    \Big(F_{n}^{\text{out}}(s)E_{n}(s)  C_{n}(s) \frac{1}{s \, c_{n}} + 1 \Big) F_{n}^{\text{in}}(s) \hat{u}_{n-1}(s).
\end{align*}
Since 
\begin{align*}
E_{n}(s) = \frac{s^2 c_{n} \gamma_{n+1}}{s^2 c_{n} \gamma_{n+1}
- K_{n}(s)(\gamma_{n+1}F_{n}^{\text{out}}(s) + c_{n}\Phi_{n+1}^{\text{in}}(s))},
\end{align*}
where $K_{n}(s) = K_{n}^p(s+1/T_{n}^I)\big/(T_{n}^F s+1)$ 
so that $C_{n}(s)=\frac{1}{s}K_{n}(s)$,
it can be seen that
\begin{align*}
    &\lim_{s\rightarrow 0}
    s \frac{1}{s \, c_{n}}
    \Big(F_{n}^{\text{out}}(s)E_{n}(s)  C_{n}(s) \frac{1}{s \, c_{n}} + 1 \Big) F_{n}^{\text{in}}(s) \nonumber\\
    &= \frac{1}{c_{n}} \Big( \frac{c_{n} \gamma_{n+1}}{-K_{n}(0)(\gamma_{n+1}+c_{n})} K_{n}(0)\frac{1}{c_{n}} + 1 \Big) \nonumber \\
    &=\frac{1}{c_{n}+\gamma_{n+1}}. 
    %\label{eq:closedloop_intgain}
\end{align*}
Therefore, with $\gamma_{n}=c_{n}+\gamma_{n+1}$,
\begin{align} \label{eq:closingtheloopC}
   \hat{h}_{n}^{\ds}(s) =\frac{1}{s \, \gamma_{n}}
   \Phi_{n}^{\text{in}}(s) \hat{u}_{n-1}(s),
\end{align}
where $\Phi_{n}^{\text{in}}(0)=1$. As such,~\eqref{eq:closingtheloopA},~\eqref{eq:closingtheloopB}, and~\eqref{eq:closingtheloopC}, continue to hold in the same form for all decrements of controlled gate index $n$, down to and including $n=\nu_m+1$. Since $\gamma_{\mu_{m}^{\ds}+1}=c_{\mu_{m}^{\ds}+1}$, once $n=\nu_m+1$, 
\[\gamma_n = c_{n} + \gamma_{n+1} = 
{\textstyle \sum_{\nu_m+1\leq \nu \leq \mu_{m}^{\ds}+1}} ~c_\nu. 
\] 
In view of~\eqref{eq:closingtheloopC}, this is the integrator gain for the target transfer function from $\hat{u}_{\nu_m}$ to $\hat{h}_{\nu_m+1}^{\ds}$, with all (if any) consecutive downstream controllers in place, as claimed.}
\end{proof}

\begin{remark} \label{remark integrator gain d to h}
Using the approach in the proof of Lemma \ref{lemma integrator gain}, it can be shown that when all control loops are closed, the integrator gain of the transfer function from each $\hat{d}_{\mc{l}}$ to $\hat{h}^{\ds}_k$ is equal to $\pm\frac{1}{\sum_{n=1}^N c_n}$ ($+$ for $d_0$ and $-$ for all of the off-takes). 
\end{remark}

\begin{theorem}\label{theorem MIMO stability}
 Given \mc{decentralized compensator design order $\bm{\nu}=(\nu_1,\ldots, \nu_{N-1})$}, and $\phi_{\text{PM}}\in(0,90)$, for each design step $m\in\{1,\ldots,N-1\}$, there exists a feedback compensator $C_{\nu_{m}}$ of the form (\ref{C_i form}) that stabilizes the SISO plant $G^{[m]}$ with a phase margin of $\phi_{\text{PM}}$ deg.
 Moreover, the recursive SISO design procedure ensures stability of the MIMO closed-loop system
 \begin{equation}
 \hat{\bm{y}}(s) = \left(\bm{I}-\bm{G}(s)\,\bm{C}(s) \right)^{-1}\bm{G}^d(s) \hat{\bm{d}}(s),
 \end{equation}
 where $\bm{C}=\operatorname{diag}(C_1,C_2,\ldots,C_{N-1})$ and, \mc{in addition}, zero steady-state error differences in response to constant disturbances; i.e., $0=\lim_{t\rightarrow \infty} \bm{y}(t) = \lim_{s \rightarrow 0} s\hat{\bm{y}}(s)$.
\end{theorem}

\begin{proof}
It is shown by induction that  at each step $m$ of the recursion, the transfer functions in (\ref{G partially closed loop}) have the following properties:
\begin{enumerate}
\item For $k,\mc{n}\in \nu_{m} ~\cup~ \overline{\bm{\nu}}^{[m]}$, and $\mc{l}\in\{1,\ldots,N+2\}$, the transfer functions $H^{[m]}_{k,\mc{n}}$ and $J^{[m]}_{k,\mc{\smash{l}}}$ have at most one integrator pole at $s=0$ and no other unstable poles;
\item For $k\in\underline{\bm{\nu}}^{[m]}$, $\mc{n}\in \nu_{m}\cup \overline{\bm{\nu}}^{[m]}$, and $\mc{l}\in\{1,\ldots, N+2\}$, the transfer functions $H^{[m]}_{k,\mc{n}}(s)$ and $J^{[m]}_{k,\smash{l}}(s)$ have a zero at $s=0$ and no unstable poles. 
\end{enumerate}
The remaining columns $\bm{H}^{[m]}_{\bullet,\underline{\bm{\nu}}^{[m]}}(s)=0$ as per (\ref{H definition}) and (\ref{Gtilde}). 

First, observe that for $m=1$, $\underline{\bm{\nu}}^{[m]}=\emptyset$, $\bm{H}^{[1]}=\bm{G}$, and $\bm{J}^{[1]}=\bm{G}^d$, where all entries have one integrator pole and no other unstable poles as discussed in Section \ref{sec frequency domain one pool}. For the induction step, suppose ${H}_{k,\mc{n}}^{[m]}$ and ${J}_{k,\smash{l}}^{[m]}$ have the required properties for given $m\in\{1,\ldots,N-2\}$. Note that $G^{[m]}$ as defined in  (\ref{G^[i] definition}) has \mc{a simple pole at $s=0$ and corresponding} non-zero integrator gain as established in Lemma \ref{lemma integrator gain}.
Let $C_{\nu_{m}}$ of the form (\ref{C_i form}) be such that $\left(1-G^{[m]}\,C_{\nu_{m}}\right)^{-1}$ is stable; the existence of such for the given structure of $G^{[m]}$ is established after the following preliminary developments. \mc{Writing the} first row of (\ref{G partially closed loop}) as
\begin{equation}
\hat{y}_{\nu_{m}} = {G}^{[m]} \hat{u}_{\nu_{m}} + \sum_{\mc{n} \in \overline{\bm{\nu}}^{[m]}} {H}^{[m]}_{\nu_{m},\mc{n}}\hat{u}_{\mc{n}} + \sum_{l=1}^{N+2} {J}_{\nu_{m},\smash{l}}^{[m]} \hat{d}_{l-1},  \label{y sigma i induction}
\end{equation}
closing the loop
% in (\ref{y sigma i induction})
with $\hat{u}_{\nu_{m}}=C_{\nu_{m}}\hat{y}_{\nu_{m}}$ gives
\begin{align}
\hat{y}_{\nu_{m}} =& \left(\! 1-G^{[m]}\,C_{\nu_{m}}\right)^{-1} \! \left(  \sum_{\mc{n} \in \overline{\bm{\nu}}^{[m]}} {H}^{[m]}_{\nu_{m},\mc{n}}\hat{u}_{n} 
\!+\! 
\sum_{l=1}^{N+2} {J}_{\nu_{m},\smash{l}}^{[m]} \hat{d}_{l-1} \!\right) \nonumber \\
=& \sum_{\mc{n} \in \overline{\bm{\nu}}^{[m]}} {H}^{[m+1]}_{\nu_{m},\mc{n}}\hat{u}_{\mc{n}} + \sum_{l=1}^{N+2} {J}_{\nu_{m},\smash{l}}^{[m+1]} \hat{d}_{l-1}. \label{y sigma i closed loop}
\end{align}
Since \mc{each of the} terms ${H}^{[m]}_{\nu_{m},\mc{n}}$ and $J_{\nu_{m},\smash{l}}^{[m]}$ have at most a single integrator, and since $\left(1-{G}^{[m]}\,C_{\nu_{m}}\right)^{-1}$ has a double zero at $s=0$ due to the integrator in ${G}^{[m]}$ and \mc{the integrator in} $C_{\nu_{m}}$, the terms ${H}^{[m+1]}_{\nu_{m},\mc{n}}$ and ${J}_{\nu_{m},\smash{l}}^{[m+1]}$ all have a zero at $s=0$. Moreover, due to stability of $\left(1-{G}^{[m]}\,C_{\nu_{m}}\right)^{-1}$, and since ${H}^{[m]}_{\nu_{m},\mc{n}}$ and ${J}_{\nu_{m},l}^{[m]}$ by hypothesis have no  unstable poles except for the integrator cancelled by $\left(1-{G}^{[m]}\,C_{\nu_{m}}\right)^{-1}$, the transfer functions $ {H}^{[m+1]}_{\nu_{m},\mc{n}}(s)$ and ${J}_{\nu_{m},\smash{l}}^{[m+1]}(s)$ have no unstable poles.  Noting that  $\nu_{m}\in\underline{\bm{\nu}}^{[m+1]}$ and $\overline{\bm{\nu}}^{[m]}=\nu_{m+1}\cup\overline{\bm{\nu}}^{[m+1]}$, it follows that $ {H}^{[m+1]}_{\nu_{m},\mc{n}}(s)$ and ${J}_{\nu_{m},\smash{l}}^{[m+1]}$ satisfy \mc{property} 2) for \mc{design step} $m+1$.

Substituting (\ref{y sigma i closed loop}) into the other rows in (\ref{G partially closed loop}), $k\neq \nu_{m}$, gives
\begin{align}
\hat{y}_{k} &=  {H}^{[m]}_{k,\nu_{m}} C_{\nu_{m}} \left(\sum_{\mc{n} \in \overline{\bm{\nu}}^{[m]}} {H}^{[m+1]}_{\nu_{m},\mc{n}}\hat{u}_{\mc{n}} + \sum_{l=1}^{N+2} {J}_{\nu_{m},\smash{l}}^{[m+1]} \hat{d}_{l-1} \right) \nonumber \\
& \quad + \sum_{\mc{n} \in \overline{\bm{\nu}}^{[m]}} {H}^{[m]}_{k,\mc{n}}\hat{u}_{\mc{n}} + \sum_{l=1}^{N+2} {J}_{k,\smash{l}}^{[m]} \hat{d}_{l-1}  \nonumber \\
&=  \sum_{\mc{n} \in \overline{\bm{\nu}}^{[m]}} {H}^{[m+1]}_{k,\mc{n}}\hat{u}_{\mc{n}} + \sum_{l=1}^{N+2} {J}_{k,\smash{l}}^{[m+1]} \hat{d}_{l-1}.  \label{y j i induction step}
\end{align}
\mc{The} two cases $k\in\overline{\bm{\nu}}^{[m]}$ and $k\in\underline{\bm{\nu}}^{[m]}$ \mc{require separate} consideration.

{\em Case $k\in\overline{\bm{\nu}}^{[m]}$.} Note that: i) ${H}^{[m]}_{k,\nu_{m}}$ has at most one integrator by hypothesis; ii) $C_{\nu_{m}}$ has one integrator by design,  iii) ${H}^{[m+1]}_{\nu_{m},\mc{n}}$ has a zero at $s=0$ as established above, and iv) none of these terms have any other unstable poles,  the product ${H}^{[m]}_{k,\nu_{m}} C_{\nu_{m}} {H}^{[m+1]}_{\nu_{m},\mc{n}}$, $\mc{n}\in \overline{\bm{\nu}}^{[m]}=\nu_{m+1}\cup\overline{\bm{\nu}}^{[m+1]}$, has at most one integrator and no other unstable poles. Likewise, ${H}^{[m]}_{k,\mc{n}}$ has  at most one integrator and no other right-half plane poles as assumed from the previous induction step. Thus, the sum of these terms, ${H}^{[m+1]}_{k,\mc{n}}={H}^{[m]}_{k,\nu_{m}} C_{\nu_{m}} {H}^{[m+1]}_{\nu_{m},\mc{n}}+{H}^{[m]}_{k,\mc{n}}$, $k\in\overline{\bm{\nu}}^{[m]}$, \mc{aligns} with property 1). The same steps can be repeated to verify that ${J}^{[m+1]}_{k,\smash{l}}$ \mc{is also consistent with property} 1).

{\em Case $k\in\underline{\bm{\nu}}^{[m]}$.} \mc{By hypothesis, for $l\in \nu_{m}\cup \overline{\bm{\nu}}^{[m]}$,} the terms ${H}^{[m]}_{k,\smash{l}}$ have a zero instead of a pole at $s=0$. The  zeros of  ${H}^{[m]}_{k,\nu_{m}}$ and ${H}^{[m+1]}_{\nu_{m},\smash{l}}$ at $s=0$ cancel the integrator in $C_{\nu_{m}}$, so that the transfer functions ${H}^{[m+1]}_{k,\smash{l}}={H}^{[m]}_{k,\nu_{m}} C_{\nu_{m}} {H}^{[m+1]}_{\nu_{m},\smash{l}}+{H}^{[m]}_{k,\smash{l}}$ also have a zero at $s=0$, and no unstable poles. The same holds for  ${J}^{[m+1]}_{k,\smash{l}}$. That is, for  $k\in\underline{\bm{\nu}}^{[m]}$, these terms all satisfy property 2).

Let $\bm{H}^{[N]}$ and $\bm{J}^{[N]}$ denote the closed-loop transfer functions \emph{after} closing the last control loop, as per~\eqref{H definition}  and~\eqref{J definition} with $\underline{\bm{\nu}}^{[N]}=\bm{\nu}$. It follows that $\bm{H}^{[N]}$ is the zero matrix, and all $J^{[N]}_{k,\smash{l}}$ satisfy property 2), i.e., stable with one zero at $s=0$.
 Then, the final value theorem implies $\lim_{t\rightarrow \infty}y_{n}(t)=0$ for all constant disturbances $d_n$, $n\in\{1,\ldots,N\}$.

It remains to establish existence of a stabilizing compensator~\eqref{C_i form} for each step $m\in\{1,\ldots,N-1\}$. Since $G^{[m]}$ and $C_{\nu_{m}}$ both have one integrator and no other unstable poles, stability of the corresponding SISO closed-loop holds if and only if the Nyquist plot (with an infinitesimal indentation of the Nyquist contour to the right of the \mc{two} poles at $s=0$) does not encircle the point $(-1,0)$ in the complex plane~\cite[Thm.~A.1.14]{curtainzwart}.  In particular, encirclement of $(-1,0)$ \mc{is} prevented by ensuring that the phase of $-G^{[m]}\,C_{\nu_{m}}$ is greater than $-180$\,deg and less than  $0$\,deg for all positive frequencies below a single cross-over frequency where the magnitude is made unity. For instance, Figure \ref{fig Nyquist} shows the Nyquist plot for the tenth step (gate $10$) of the example channel. For positive $\omega$ below crossover, the Nyquist plot stays in the third quadrant since the phase of $\mc{-}{G}^{[10]}C_{10}$ is greater than $-180$ and less than $-90$ degrees. At higher frequencies where the phase  is outside $(-180,-90)$\,deg, the magnitude is well below unity so that $(-1,0)$ is not encircled. The Nyquist plot is closed in a clockwise direction as the Nyquist contour traverses the indentation around $s=0$.

\begin{figure}[tbp!]\centering 
\includegraphics[width=.7\columnwidth]{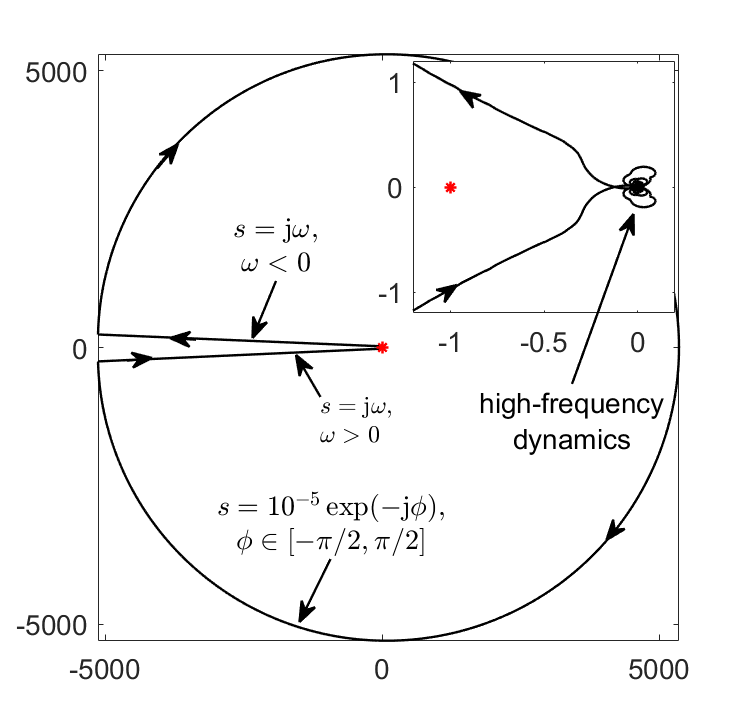}
\caption{Nyquist plot of $\mc{-}{G}^{[10]}C_{10}$ from Figure \ref{fig shaped GK}. }
\label{fig Nyquist}
\end{figure}

\mc{For the purpose of establishing the feasibility of closed-loop stability,} the low-pass filter time constant $T^F_{\nu_{m}}$ can be set to an arbitrary value. Let $\i=\sqrt{-1}$. Due to the integrator in ${G}^{[m]}$ and since $|(1+\i\omega\,T^F_{\nu_{m}})^{-1}|$ is strictly decreasing, there exists $\omega_0>0$ such that given any $\omega\leq \omega_0$,
\begin{equation}
\left| \frac{{G}^{[m]}(\i\varphi)}{1+\i\varphi\, T^F_{\nu_{m}}} \right| < \left| \frac{{G}^{[m]}(\i\omega)}{1+\i\omega\,T^F_{\nu_{m}}} \right| \quad \text{for all } \varphi > \omega. 
\end{equation} 
 Thus, if the  crossover frequency, $\omega^c_{\nu_{m}}$, of the $m$-th local loop is chosen such that    $\omega^c_{\nu_{m}}\leq \omega_0$, then $| {G}^{[m]}(\i\omega)/(1+\i\omega\,T^F_{\nu_{m}})|$ crosses 0dB only once. Due to the  positive integrator gain of ${G}^{[m]}$  as given in (\ref{integrator step i}), the phase of  $\angle {G}^{[m]}(\i\omega)$ is arbitrarily close to $-90$\,deg for sufficiently small $\omega$. See also Figure \ref{fig open loop G}. Therefore, $\omega^c_{\nu_m}\leq \omega_0$ can be chosen so that additionally
\begin{equation}
\angle\left( \frac{ {G}^{[m]}(\i\omega^c_{\nu_{m}})}{1+\i\omega^c_{\nu_{m}}\,T^F_{\nu_{m}}}\right)\geq -180+\phi_{\text{PM}}+0.5(90-\phi_{\text{PM}}). \label{phase margin 1}
\end{equation} 
Then, $T^I_{\nu_{m}}$ can be chosen such that 
\begin{equation}
\angle\left( \frac{\i\,\omega^c_{\nu_{m}}\,T^I_{\nu_{m}}+1}{\i\,\omega^c_{\nu_{m}}\,T^I_{\nu_{m}}}\right) = 0.5(\phi_{\text{PM}}-90). \label{phase margin 2}
\end{equation}
Conditions (\ref{phase margin 1}) and (\ref{phase margin 2}) ensure that the phase of $\mc{-}{G}^{[m]}C_{\nu_{m}}$ at crossover is greater than $-180+\phi_{\text{PM}}$, i.e., the phase margin specification is satisfied. Finally, $K^p_{\nu_{m}}$ can be chosen such that 
\begin{equation}
\left|{G}^{[m]}(\i\omega^c_{\nu_{m}})C_{\nu_{m}}(\i\omega^c_{\nu_{m}})\right|=1.
\end{equation}

To summarize, the SISO loop-shaping problem at each step is well posed, in  the sense that decentralized controllers satisfying the phase margin condition can be found. Moreover, designing the decentralized controllers via the $N-1$ sequential SISO compensator parameter tuning steps ensures the overall MIMO system is robustly stable, with the desired steady-state disturbance rejection property.
\end{proof}

\begin{remark}
 The integrator gain in (\ref{integrator step i}) obviously decreases for larger sums in both denominators.  From the last part of the proof of Theorem \ref{theorem MIMO stability}, it can be seen  that  the loop-shaping design criteria can be satisfied if crossover is placed in the frequency range where the integrator dynamics dominate. While experience  suggests that in many cases a  faster crossover can be achieved (see, e.g., Figure \ref{fig shaped GK}), in some cases small integrator gain poses a limitation on the achievable bandwidth. Therefore, from a performance perspective it is desirable to choose the design order $\bm{\nu}$ such that the integrator gain of $G^{[m]}$ remains (relatively) large. By (\ref{integrator step i}), this can be achieved by designing the controllers in a ``contiguous'' manner, i.e., starting with one (arbitrary) gate, and then designing the other controllers for gates directly next to control loops that have already been closed in previous steps. That way, there is only one summand in one of the two denominators in (\ref{integrator step i}). For $\bm{\nu}=(2,1,3)$, in the case of four pools for instance, controller 2 balances pools 2 and 3, then controller 1 can make pool 1 follow the relatively slower, already balanced pools 2 and 3, and finally controller 3 can make pool 4 follow balanced pools 1-3. By contrast, for $\bm{\nu}=(1,3,2)$, controller 2 would need to balance the relatively slow, balanced pools 1-2 with the equally slow combined pools 3-4. See also the bottom example in Figure \ref{fig recursive loop-shaping 2}.
\end{remark}

\subsection{Feedforward of measured disturbances}\label{sec:feedforward}
In order to improve transient performance, the balancing controllers can be augmented with feedforward terms. In particular, the top inflow, $d_0$, and the  outflow over the bottom gate, $d_{N+1}$, are assumed to be measured, while all other outflows due to off-takes in the pools are unmeasured. This is a common situation in practice. The control inputs \mc{become}
\begin{equation}
\hat{u}_n(s) = C_n(s) \hat{y}_n(s) + \mc{D}^{\text{in}}_n \hat{d}_0(s)  + \mc{D}^{\text{out}}_n \hat{d}_{N+1}(s). \label{eq:feedforward}
\end{equation}
Large step-like changes in the top inflow can potentially overwhelm the top pool in the channel before the feedback controllers can react. The purpose of these feedforward terms is to quickly distribute such large changes in the flow into (or out of) the balancing pools over multiple pools. The feedforward gains are set to $\mc{D}^{\text{in}}_n=0.85^n$ and $\mc{D}^{\text{out}}_n=0.85^{N-n}$, $\mc{n\in\{}1,\ldots,N-1\}$. That is, the first pool effectively receives only 15\% of an inflow step change, the second pool 15\% of 85\%, and so on.  In principle, the feedforward gains could be scaled to account for different lengths and storage capacities of the pools, but this was found to make only a minor difference. Similar feedforward terms could be added if the other outflows $d_n$, $n=1,\ldots,N$, are measured.

\section{Numerical simulations } \label{sec:simulations}

The \mc{sequential approach to balancing controller} design is \mc{first} verified in numerical simulations \mc{for} the channel consisting of the pools with dimensions as given in Figure \ref{fig channel data}; \mc{physical field trial results on a different channel are presented in Section~\ref{sec:fieldtest}}. The \mc{decentralized} balancing controllers are \mc{applied to} the first 20 pools of the channel, while the remaining \mc{13} pools \mc{operate under} distant-downstream control; see, e.g.,~\cite{cantoni2007control,cantonimareels2021encyclopedia} for discussion on the design of such distant-downstream controllers. This represents a realistic configuration where the channel is operated in an overall demand-driven fashion, with the balancing controllers \mc{used to create} a storage buffer in the upper part of the channel \mc{for reducing sensitivity to mismatch between the supply to and demand from the top part of the channel. The related use of balancing controllers to mitigate string instability is explored in~\cite{strecker2025ccta}.}

 \begin{figure}[tbp!]\centering 
\includegraphics[width=1\columnwidth]{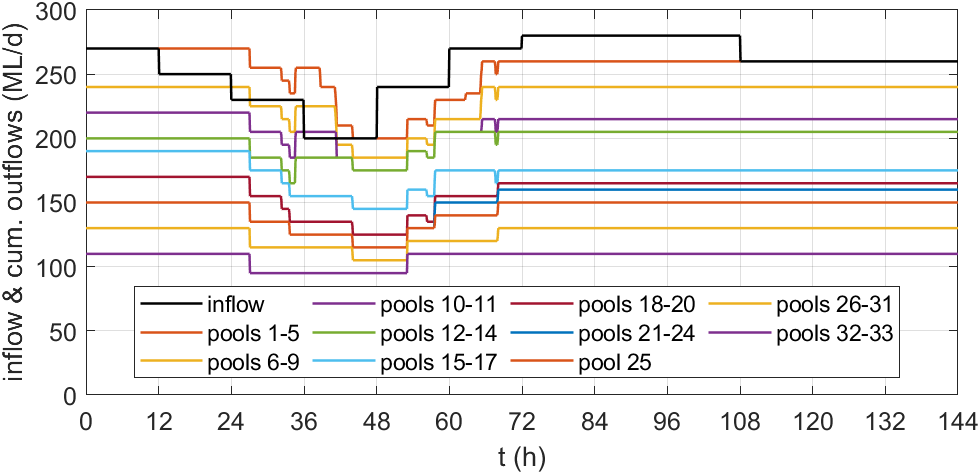}
\caption{Inflow and cumulative outflows (sum of all off-takes in or downstream of each pool) used in simulations throughout this paper.}
\label{fig in/outflows}
\end{figure}

The dynamics in each pool are represented by the full nonlinear Saint Venant Equations (\ref{StVenant1})-(\ref{StVenant2}), which for the simulations are semi-discretized in space using a finite difference scheme (``method of lines''). The resulting system of high-order ordinary differential equations (ODEs) are then coupled with the ODEs of the \mc{decentralized PI} controller dynamics, and simulated using standard ODE solvers in Matlab. Throughout this section, the inflow and outflows given in Figure \ref{fig in/outflows} are used. That is, there is under-supply (inflow less that cumulative outflows from pool 1) during the time period between 12 and 48 hours, and over-supply between 48 and 108 hours, after which the inflow matches the outflows. The outflows vary relatively frequently between 24 and 72 hours. Beyond that, the outflows are kept constant so that the system can settle in a way that makes the \mc{impact of} demand-supply mismatch clearly visible. 

The decentralized controllers are designed in the order $\bm{\nu}=(1,2,3,\ldots,19)$, with a 50\,deg phase margin for each local SISO loop. Unless noted otherwise, the feedback controllers are augmented with top and bottom flow feedforward terms as described in Section \ref{sec:feedforward}.

\subsection{Simulation results for $N=20$}

 \begin{figure}[htbp!]\centering 
\includegraphics[width=.99\columnwidth]{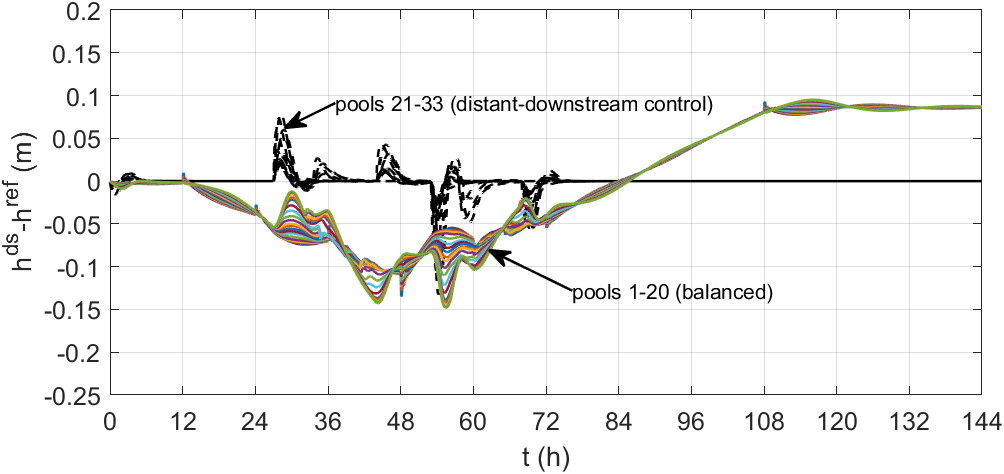}\\
\includegraphics[width=.99\columnwidth]{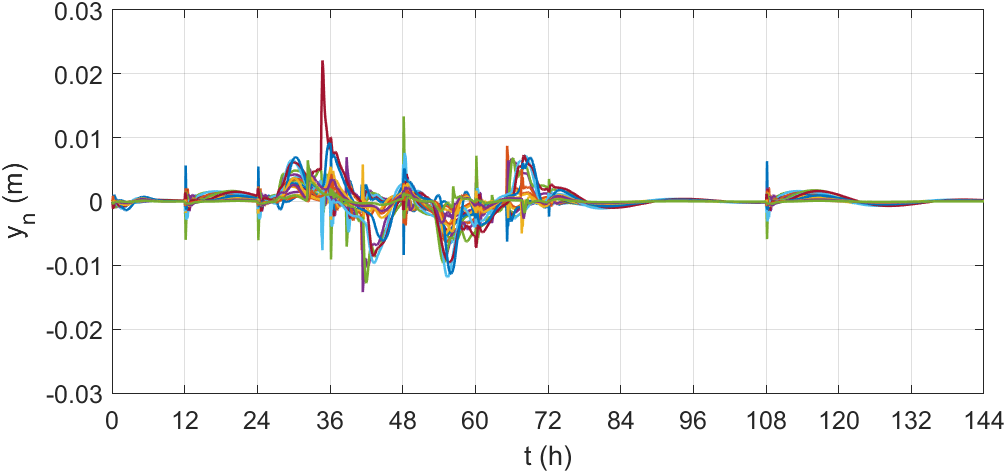}\\
\includegraphics[width=.99\columnwidth]{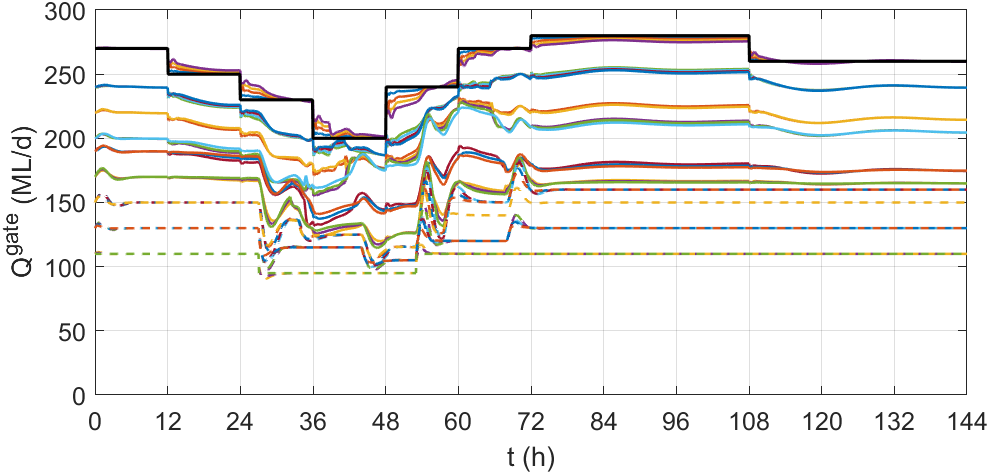}
\caption{Simulated water level errors ($\hds_n-\hrref_n$), difference in level errors between consecutive pools ($y_n$), and gate flows ($u_n$). Each solid line represents the trajectories of one water level, $y_n$, or gate flow, for the first 20 pools and 19 gates that are operated in balancing control. The dashed lines represent the remaining pools and gates operated in distant-downstream control. In the bottom plot, the solid black line represents the inflow $d_0$.}
\label{fig simulation decentralized 20}
\end{figure}

The simulated water level deviations and flows are shown in Figure \ref{fig simulation decentralized 20}.  It can be seen how the water levels of the first $N=20$ pools can drop below their reference values in a coordinated manner during periods of under-supply, and rise above the reference during over-supply. The outflow changes cause further fluctuations in the levels and control inputs. Once the inflow matches the outflows, the water levels settle approximately 8\,cm above the reference, due to the excess water stored in the channel from the previous period of over-supply.  The difference in tracking errors between consecutive pools, $y_n$, generally stay very small, and short spikes due to step changes in the outtakes are quickly corrected by the feedback controller. Pool 14 is relatively short but has a relatively large outtake, which causes the relatively large spike in $y_{14}$ at $t=35$ hours. In the gate flow plot, it can be seen for instance that  after times 12 and 24 hours, the flow over gates 1-4 exceeds the inflow at the top (see the four coloured line above the black line). At these times, the levels in those pools drop due to under-supply, and the balancing controllers distribute this under-supply by passing more flow out of each pool than the flows in.   Some moderate oscillations are visible, most obviously after 108 hours. Such oscillatory transients are characteristic of integral action in consensus-type control~\cite{tegling2018thesis}.

\begin{figure}[htbp!]\centering 
\includegraphics[width=.99\columnwidth]{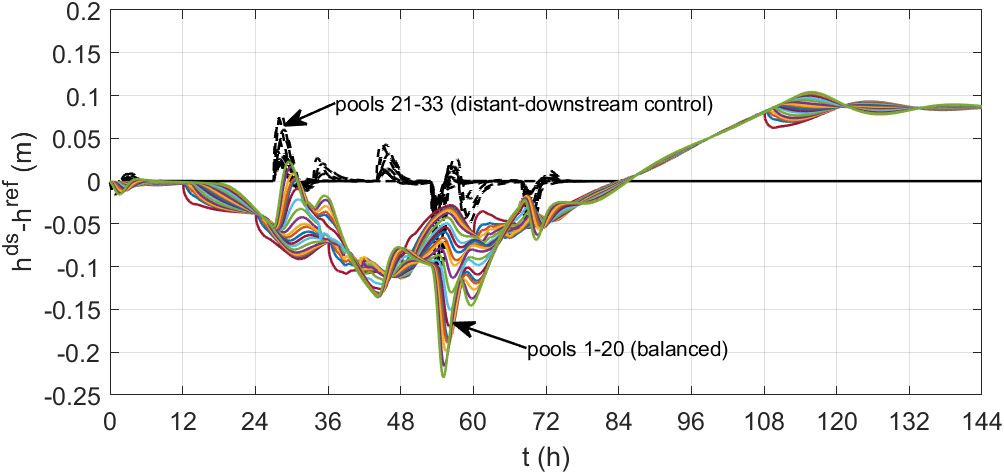}\\
\includegraphics[width=.99\columnwidth]{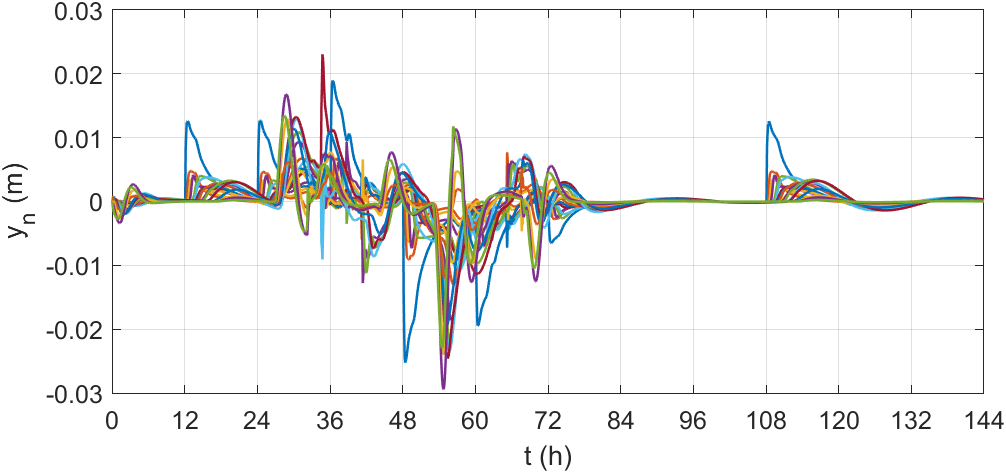}\\
\includegraphics[width=.99\columnwidth]{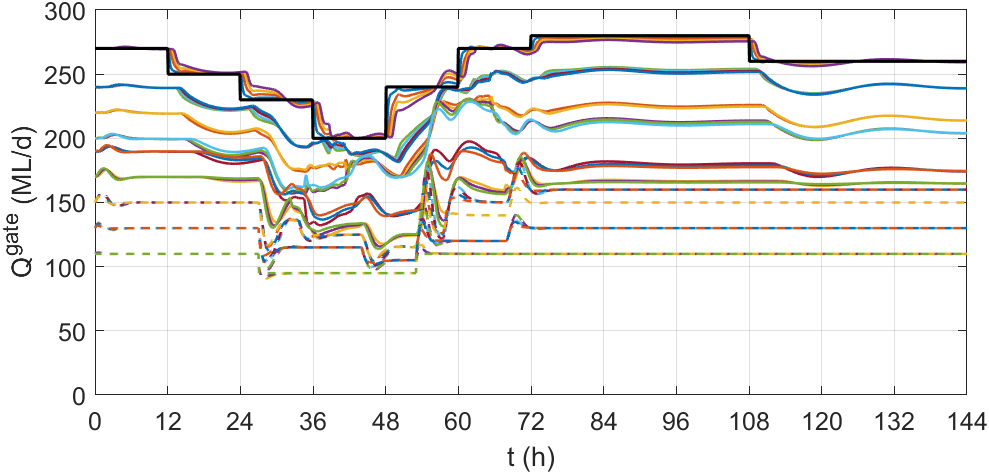}
\caption{{ Simulation results if no feedforward is used, with the first $N=20$ pools in balancing control.}}
\label{fig simulation decentralized 20 no FF}
\end{figure}

For comparison, simulation results for the case with pure feedback control, i.e., without the feedforward terms from Section \ref{sec:feedforward}, are shown in Figure \ref{fig simulation decentralized 20 no FF}. Without pre-emptive adjustments in the control inputs, step changes in the disturbances cause larger transient tracking error differences between consecutive pools, $y_n$, and consequently larger level tracking errors. Nevertheless, the  controllers succeed in balancing  the water level tracking errors over time.

\subsection{Sensitivity to number of pools}
 \begin{figure}[htbp!]\centering 
\includegraphics[width=.99\columnwidth]{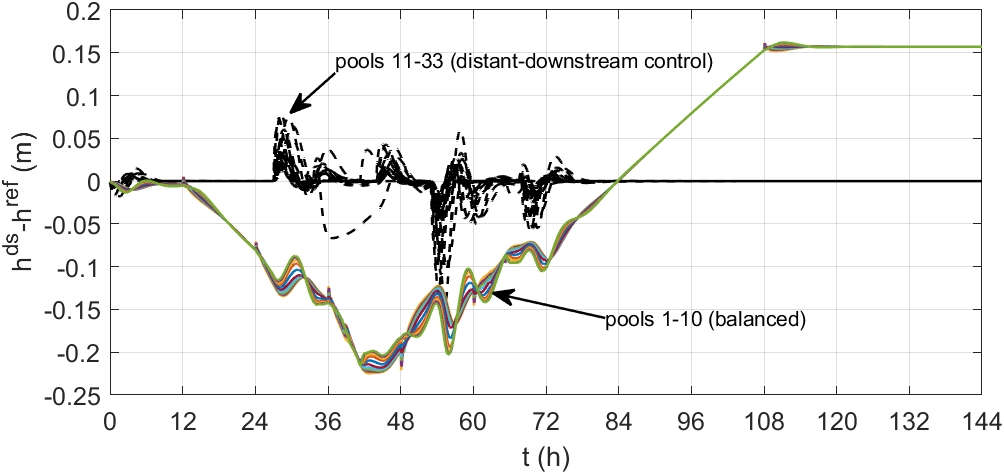}\\
\includegraphics[width=.99\columnwidth]{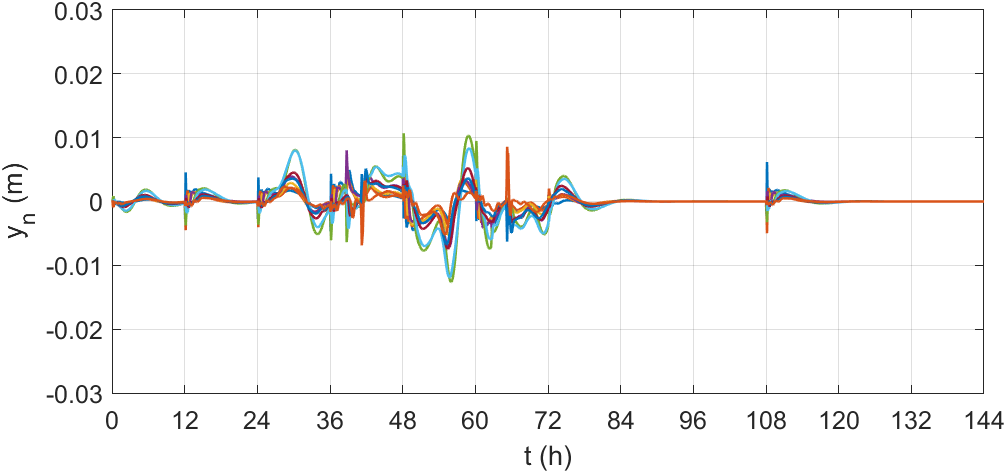}\\
\caption{{ Simulated water level deviations with only the first  $N=10$ pools in balancing control.}}
\label{fig simulation decentralized 10}
\end{figure}
As indicated in Figure \ref{fig dV to dh }, the number $N$ of balanced pools and their combined storage capacity affects the sensitivity of the water level errors to flow mismatch.
Figure \ref{fig simulation decentralized 10} shows simulations similar to Figure \ref{fig simulation decentralized 20} but with only the first $N=10$ pools in balancing control. That is, the total storage capacity in the balanced pools is smaller than for $N=20$. Consequently, the water level deviations are almost doubled compared to Figure \ref{fig simulation decentralized 20}.  The oscillations after 108 hours are less than in the case with $N=20$.  In general, for smaller $N$ the control problem tends to become ``easier'' in the sense that the water level deviations can be kept tighter together, i.e.,  $\max_{n,m}|(\hrref_{n} - \hat{h}_{n}^{\ds}(t)) - (\hrref_{m}- \hat{h}_{m}^{\ds}(t))|$ tends to be smaller, and the oscillations are less significant. This comes at the cost of reduced storage volume leading to larger deviations of the average tracking errors.

\subsection{Weighted tracking}
 \begin{figure}[tbp!]\centering 
\includegraphics[width=.99\columnwidth]{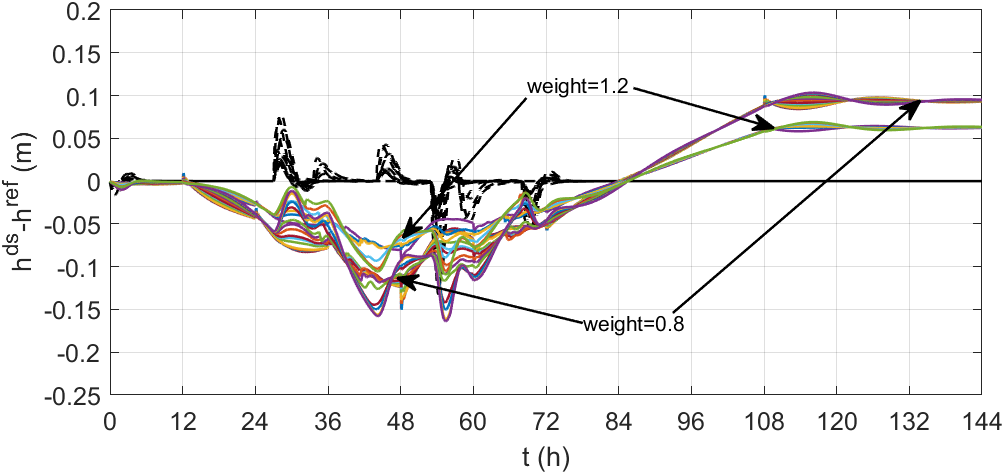}
\caption{ Simulations with weighted $y_n$ as in (\ref{y definition weighted}).}
\label{fig simulation weighted}
\end{figure}

The previous control design and simulations focused on the case where equal water level tracking performance is desired in all pools. In practice, it is sometimes desirable to have tighter control in some pools (e.g., in pools with active outtakes) compared to others. This can be addressed by weighting the water-level errors by a weight $\bm{w}=(w_1,\ldots,w_N)$ as in
\begin{align}
y_n(t) &= w_n\left(\hrref_{n} - h_{n}^{\ds}(t)\right) - w_{n+1}\left(\hrref_{n+1}- h_{n+1}^{\ds}(t)\right). \label{y definition weighted} 
\end{align}
If $y_n=0$, $\left(\hrref_{n} - h_{n}^{\ds}(t)\right)= \frac{ w_{n+1}}{w_n}\left(\hrref_{n+1}- h_{n+1}^{\ds}(t)\right)$. That is, larger $w_n$ promotes tighter water level control. The transfer function $G$ in (\ref{y MIMO}) becomes
\begin{align}
G_{n,n} &=w_{n+1}G^{\text{in}}_{n+1} - w_{n}G^{\text{out}}_{n},~ n \in\mc{\{}1,\ldots,N-1\mc{\}}, \\
G_{n,n+1} &= w_{n+1}G^{\text{out}}_{n+1},  \qquad n \in\mc{\{}1,\ldots,N-2\mc{\}}, \\
G_{n,n-1} &= -w_{n}G^{\text{in}}_{n},  \qquad n \in\mc{\{}2,\ldots,N-1\mc{\}}.
\end{align}
Then the \mc{sequential} loop-shaping \mc{method} can be executed exactly as in Section \ref{subsection recursion}. 

The resulting trajectories with weighting are shown in Figure \ref{fig simulation weighted}. Here, the weight is set to $w_n=1.2$ for $n\in\{5, 9, 11, 14, 17, 20\}$ (pools with active outtakes), and $w_n=0.8$ for all other pools. That is, level tracking in pools with outtakes is prioritized by a factor of 1.5 relative to the other pools. The tighter level control in these pools is clearly visible in Figure \ref{fig simulation weighted}.

\begin{figure}[tbp!]\centering 
\includegraphics[width=.99\columnwidth]{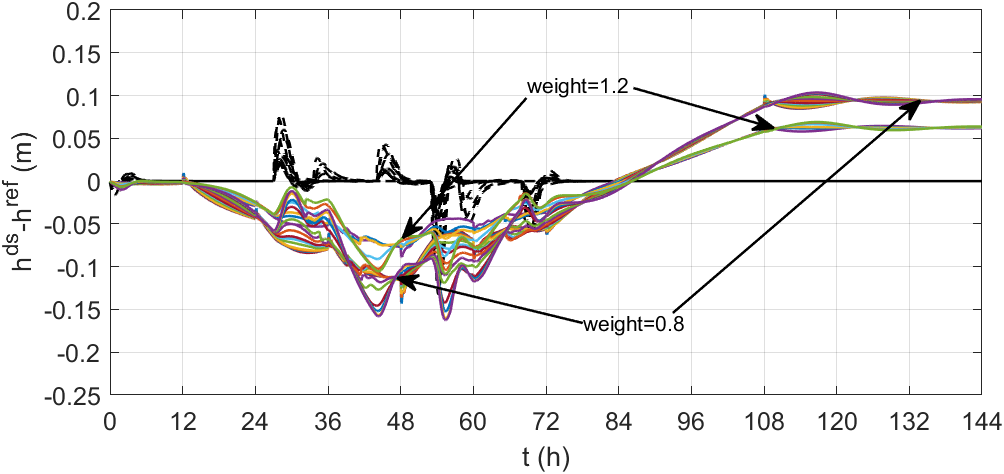}
\caption{ Simulations with weighted $y_n$ as in (\ref{y definition weighted}) but using the same controllers as in Figure \ref{fig simulation decentralized 20} designed with uniform weights $w_n=1$.}
\label{fig simulation weighted unweighted C}
\end{figure}

The \mc{relative priority of pools} can change, for instance due to outtakes becoming active. A switched control strategy, where the controllers are re-tuned every time priority weights \mc{are updated}, would be impractical in most cases. Alternatively, controllers designed  for uniform weights can be implemented with the weighted outputs (\ref{y definition weighted}). The mismatch between the plants for which the controllers are designed and the plant on which they are implemented can be handled by designing the controllers with sufficient robustness margins. Then, closed-loop stability can be verified a posteriori \mc{for the modified weights}. The trajectories for the case where $y$ is evaluated according to the same weights as in Figure \ref{fig simulation weighted}, but using the same (unweighted) controllers as in Figure \ref{fig simulation decentralized 20}, are shown in Figure \ref{fig simulation weighted unweighted C}. Overall, the difference in performance is marginal, suggesting that the original unweighted design is sufficiently robust.

\section{\red Field trial results}
\label{sec:fieldtest}

\begin{figure}[tbp!]\centering 
\includegraphics[width=.99\columnwidth]{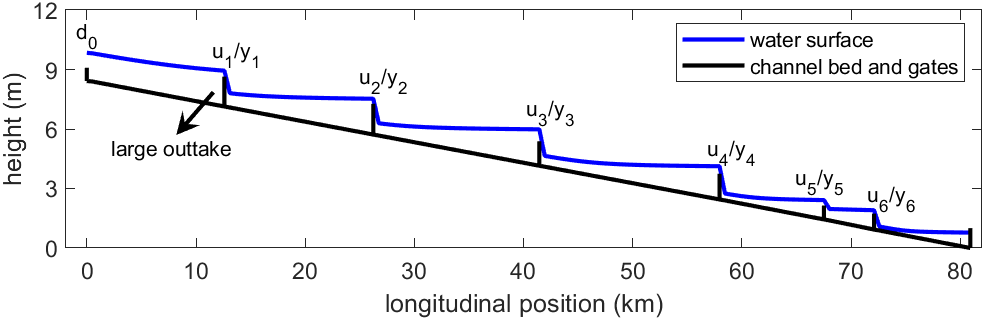}
\caption{ \red Sideview of the Waranga Western Channel  - Boort section. There are multiple outtakes in most pools but there is one outtake in the first pool that is by far the largest one. }
\label{trial side view}
\end{figure}

{\red
\mc{Decentralized balancing controllers designed according to the method proposed} in this paper \mc{were deployed} on the Boort section of the Waranga Western Channel  in Victoria, Australia, for a seven-day period in May 2025. As shown in Figure \ref{trial side view}, this channel consist of seven long pools, ranging from 4.5 to 16\,km in length, with only one pool shorter than 8.8\,km. The bed width varies between approximately 9\,m and 13\,m, and the reference   downstream water levels vary between 0.8\,m and 1.9\,m. \red The pools are relatively flat with a bottom slope of $10^{-4}$, but due to their length, the drop in elevation over each pool is so large that the water backs up only part-way up the channel. Therefore, only part of each pool acts as a storage, whereas the upstream part of each pool merely conveys the flow. \mc{The biggest (by far)} outtakes occur in the first pool. Consequently, a much higher water level is required in \mc{the first} pool for conveyance, and as such, the storage capacity is reduced somewhat compared to the other pools of similar length.

The inflow at the top, $d_0$, is set manually and can be changed only a few times per day due to hardware limitations. There are  further limitations on the inflow since water needs to be ordered from a river multiple days in advance. Consequently, there is often a large mismatch between the supply to and the demand in the channel, relative to the storage capacities of individual pools. This limits the applicability of standard decentralized distant downstream control. 

This channel is \mc{usually} managed under a hybrid combination of manual gate positioning to set nominal flow rates, and distant-downstream feedback control of the second pool. This requires frequent interventions by the operators, and often leads to large water level deviations from supply level, especially in the most upstream pool. \mc{Under} manual control, \mc{coordinated utilization of the storage capacity} in all pools is challenging. \mc{Frequently, water levels are too high in some pools, and too low in others, giving rise to inequitable quality of service along the channel.}

\begin{figure}[tbp!]\centering 
\includegraphics[width=.99\columnwidth]{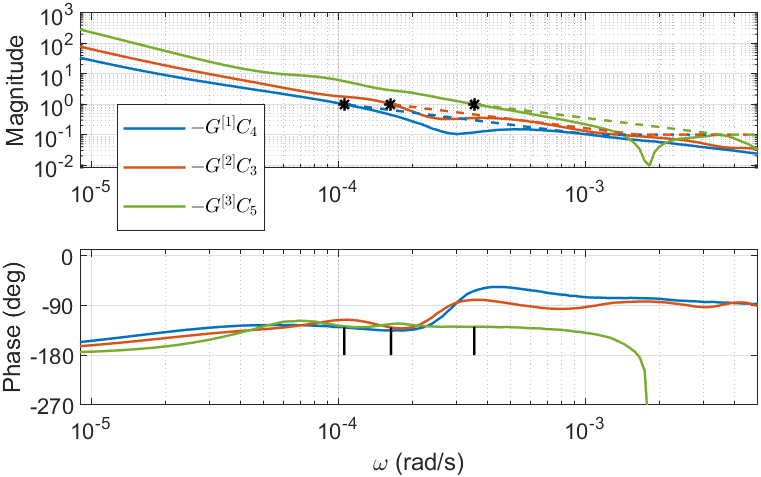}
\includegraphics[width=.99\columnwidth]{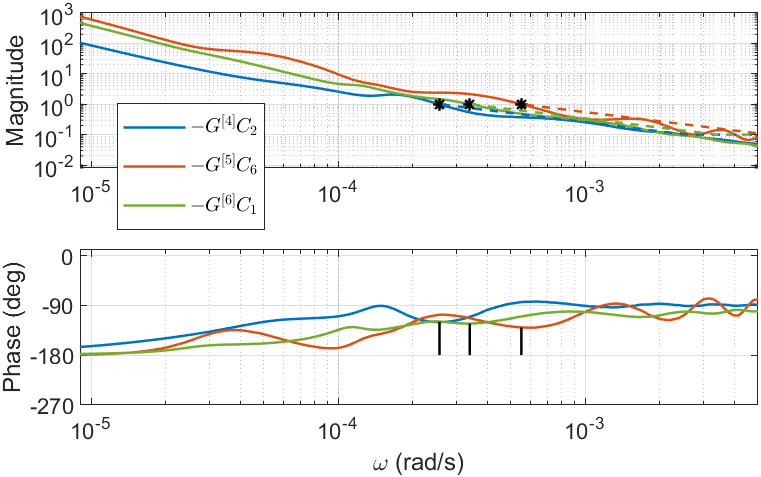}
\caption{ \red Bode plots of the shaped loops $\mc{-}{G}^{[m]}C_{\nu_{m}}$ used in the field trial, which have been designed in the order  $\bm{\nu}=(4,3,5,2,6,1)$. The dashed lines indicate the bound on the roll-off, and the black asterisks indicate crossover.}
\label{Bode trial}
\end{figure}

The \mc{decentralized balancing} controllers trialled were designed in the order $\bm{\nu}=(4,3,5,2,6,1)$ \mc{with} the weights \mc{set as} $\bm{w}=(0.9,1,1,1.25,1.43,1.11,0.7)$ in~\eqref{y definition weighted}. The weights reflect the available freeboard and prioritization of certain pools.  The transfer functions of the shaped loops are shown in Figure \ref{Bode trial}. Due to a ``notch'' in the magnitude of the transfer function from $\Qgate_4$ to $y_4$ (without any control loops closed), it was found \mc{to be} beneficial to start  designing the balancing controllers \mc{from} the fourth gate \mc{along the stretch} in balancing control, and continue ``outwards''. The crossover frequency achieved for ${G}^{[1]}C_{4}$ in Figure \ref{Bode trial} is less than for the other gates, but the achievable crossover for that gate was even worse for different design orders $\bm{\nu}$.
}

\begin{figure}[tbp!]\centering 
\includegraphics[width=.99\columnwidth]{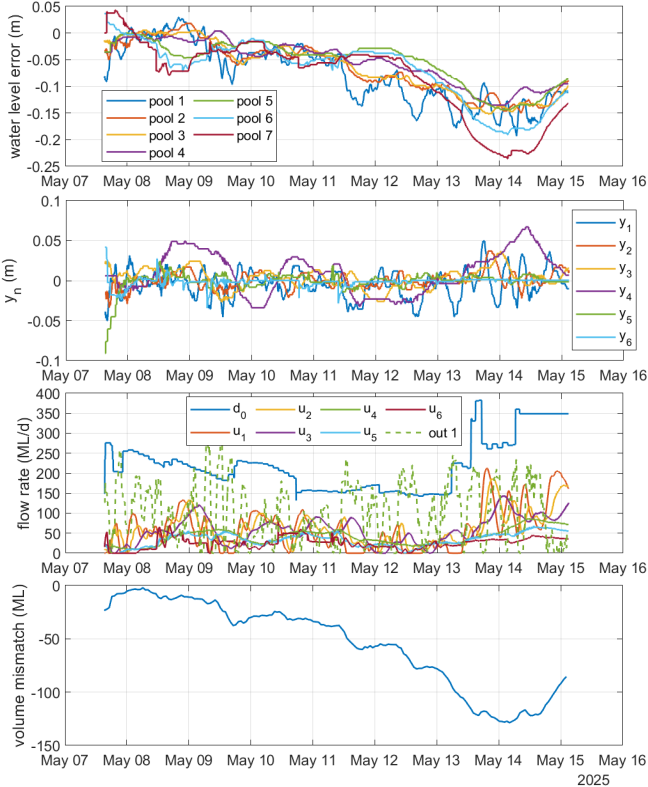}
\caption{ \red Water level errors, \mc{level error differences $y_n$,} \mc{gate flows}, and total volume mismatch in the Waranga Western Channel during the May 2025 trial period. ``out 1'' in the legend refers to the largest outtake in the first pool, but there are other unmeasured outtakes in that pool. }
\label{trajectories trial}
\end{figure}

{\red
The water-level \mc{error, gate flow}, and \mc{neighbouring} error difference trajectories observed during the trial are shown in Figure \ref{trajectories trial}. Overall, the controllers succeeded in balancing the level \mc{errors}, with all generally tracking in the same direction, and in maintaining the water-level error differences $y_n$ within a $\pm 5$ centimeter band of zero. The level in the 5th pool was controlled relatively tightly, while the 7th pool had the largest level fluctuations, in line with the setting of the weights. As expected, given the relatively small crossover achieved for the 4th gate seen in Figure \ref{Bode trial}, the corresponding level error difference $y_4$ tends to be larger than for the other gates. The bottom plot of Figure \ref{trajectories trial} reflects the accumulation of under supply over the trial period, leading to a drop in all water levels. The extent of the under supply during the trial is large because the channel was \mc{being drawn down for} the end of the season. The volume mismatch shown is the water volume in the channel relative to the volume stored with all water levels at set-point. Since this channel is much wider than deep, the width of the water surface bears little dependence on the water level. As such, the volume mismatch can be approximated with high accuracy by $\sum_{n=1}^7c_n(\hds_n-h^{\text{ref}}_n)$, where $1/c_n$ is the integrator gain for pool $n$. The large and frequent fluctuations in the first pool outtake caused the relatively high-frequency fluctuations in the level there. These were only partially propagated to the other pools by the balancing controllers. Saturation can be seen in some instances, which was handled by \mc{a standard} anti-windup \mc{scheme for PI compensators~\cite{astrom2006pid}.}
}

\begin{figure}[tbp!]\centering 
\includegraphics[width=.99\columnwidth]{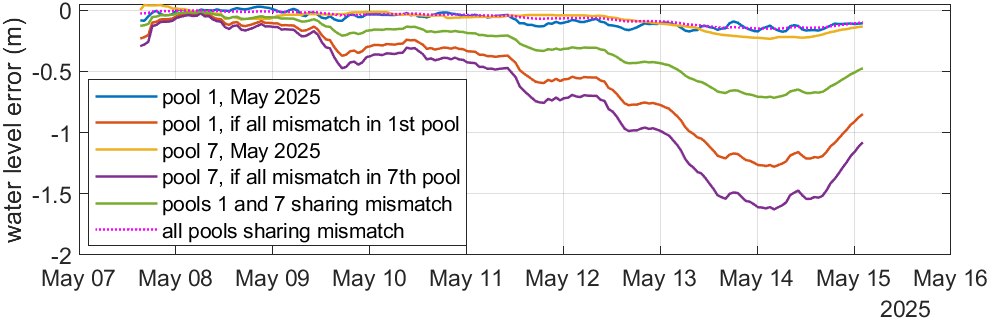}
\caption{ \red Measured water levels in the first and the most downstream pool during the trial period, compared
to the following three hypothetical scenarios: \mc{1)} mismatch concentrated in only one of these pools (similar to what would occur under distant-downstream or upstream level control, respectively); \mc{2}) mismatch distributed only among these 2 pools, but not the other 5; and \mc{3)} mismatch distributed equally among all pools. }
\label{hypothetical trial}
\end{figure}

{\red In Figure \ref{hypothetical trial}, the water levels achieved during the trial period are compared to hypothetical scenarios where all \mc{of} the demand-supply mismatch is borne by the most upstream pool (as would occur under distant-downstream control), or the most downstream pool (as with upstream-level control), or by these two pools combined. These hypothetical trajectories are obtained by dividing the volume mismatch by the respective pool capacity $c_n$. Clearly, individual pools would be unable to absorb such a large under supply.}

\begin{figure*}[htbp!]
\includegraphics[width=.48\textwidth]{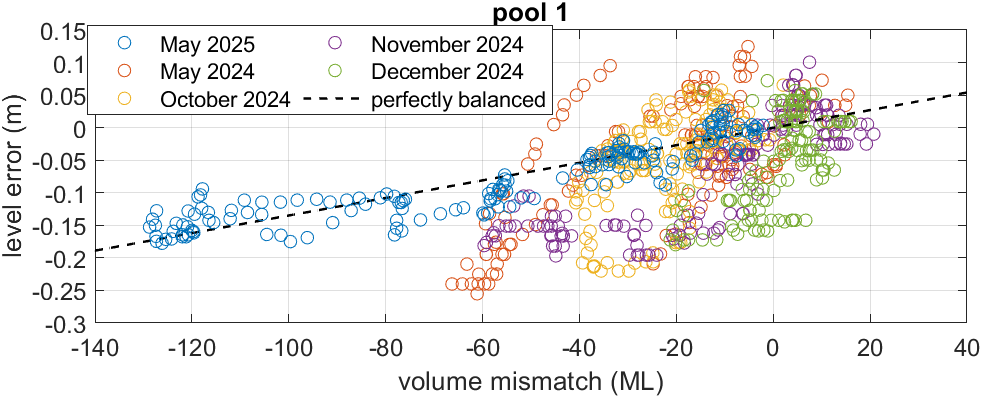}
\includegraphics[width=.48\textwidth]{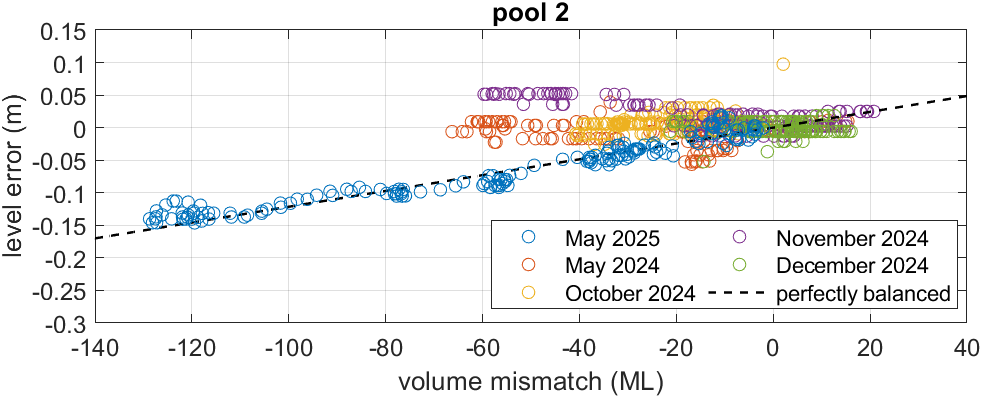}
\includegraphics[width=.48\textwidth]{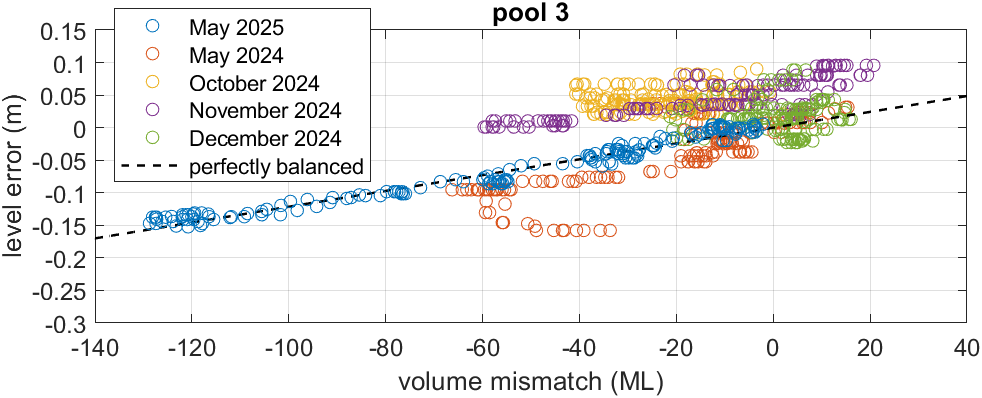}
\includegraphics[width=.48\textwidth]{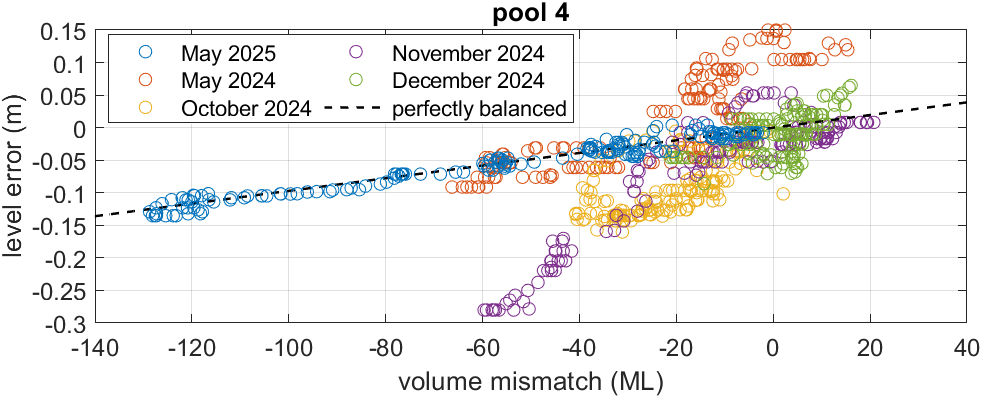}
\includegraphics[width=.48\textwidth]{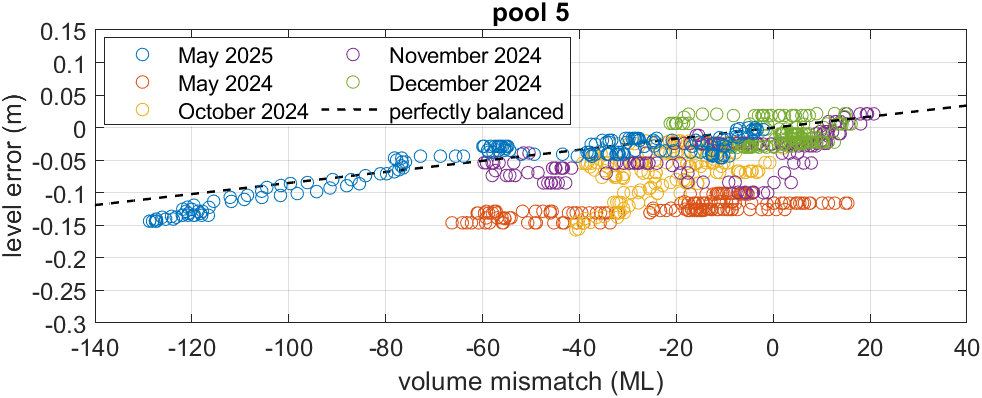}
\includegraphics[width=.48\textwidth]{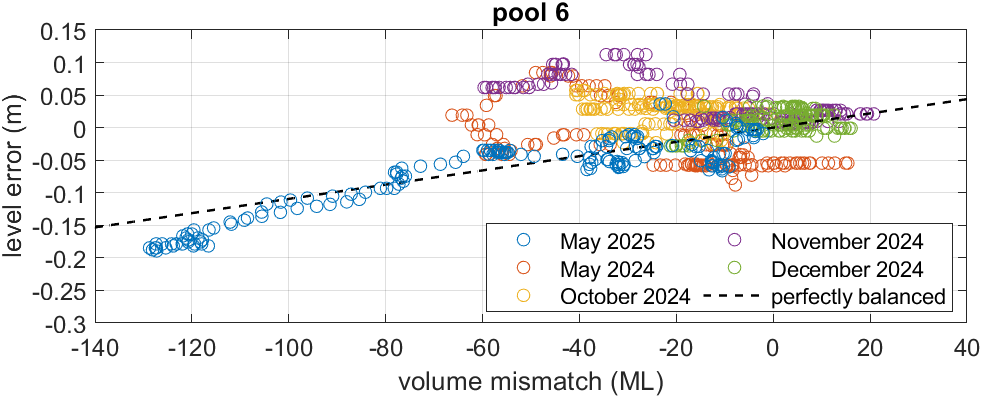}
\includegraphics[width=.48\textwidth]{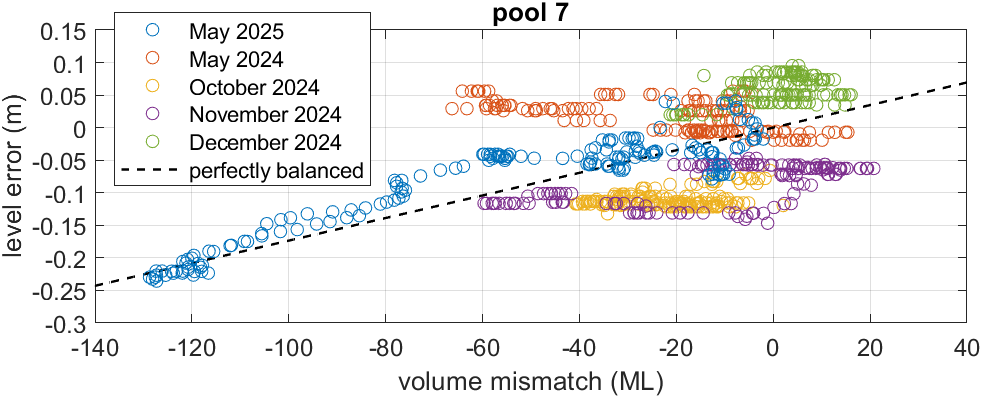}\hfill 
\includegraphics[width=.45\textwidth]{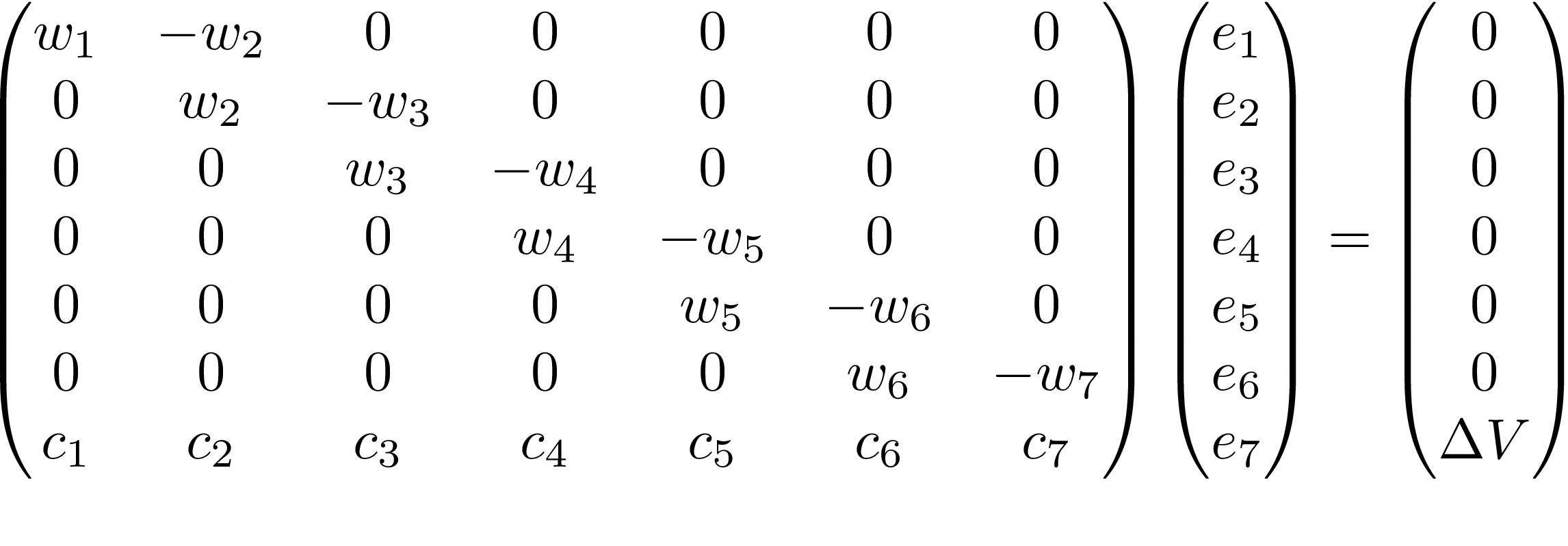}
\caption{ \red Scatter plot of the water level errors \mc{versus} the total volume mismatch at 1-hour sample \mc{intervals} during the trial period, and during three other 7-day periods under the previous mode of operation. The black line indicating perfectly balanced levels varies between the pools due to the different weights $w_n$.  This relationship between the volume mismatch, $\Delta V$, and the water level errors, $e_n$, is obtained via the equation at the bottom right. The first 6 rows of this equation are obtained by setting $y_n=0$ for $n=1,\ldots,6$, i.e., assuming perfect weighted balancing, while the last row is the linear approximation of the relationship between  level errors and  volume mismatch.   }
\label{scatter trial}
\end{figure*}

{\red For further comparison, Figure \ref{scatter trial} shows scatter plots of the level errors \mc{versus} the volume mismatch during the May 2025 trial period, and during three other 7-day periods \mc{under manual control}. By inspection, the levels during the trial period were much closer to the ideal line where all pools are balanced according to their weights. During the  periods under the previous mode of operation, there were several instances where the levels in individual pools were significantly higher than desired even though there was overall an under supply in the channel. The total extent of under supply was up to twice as much during the May 2025 \mc{trial}, yet the worst-case level drop in, e.g., the first and fourth pools, was less severe than in the comparison data.}

\section{Conclusions} \label{sec: conclusions}
A sequential \mc{design procedure} for decentralized controllers that \mc{balance} water-level deviations \mc{from setpoint is developed for} irrigation channels \mc{subject to system level supply-demand mismatch exceeding the capacity of individual pools}. By distributing mismatch \mc{among} multiple pools, water-level deviations can be \mc{moderated} without resorting to supply driven operations \mc{and} corresponding reduction in
\mc{downstream} efficiency or service depending on the sign of the mismatch. This \mc{enables} an overall demand-driven operation of channels for which traditional approaches like distant downstream control would be infeasible due to the insufficient storage capacity in individual pools. The design offers flexibility by allowing the weighting of water level errors in individual pools. 

The decentralized implementation imposes  limitations  on the achievable  coherence over the channel, i.e., on how close the water level deviations of all pools can be kept together.  For instance, some oscillations are visible in the simulations. Such limitations are common for the given model and controller structure  \cite{tegling2018thesis,bamieh2012coherence}.  This will be improved in future work by adding a supervisory control layer \mc{operating} on a slower time scale with \mc{supply and demand forecasts}. Such a supervisory layer is already commonly used to improve performance in  distant downstream control~\cite{nasir2021tcst}.

While the recursive loop-shaping approach is motivated by the irrigation channel control problem, it is not inherently limited to this application. The proofs of Lemma \ref{lemma integrator gain} and Theorem \ref{theorem MIMO stability}, for instance, only assume that the uncontrolled subsystems are proper and have one integrator. Each step involves a classical SISO loopshaping problem. The approach is likely applicable in other contexts.

\appendices
\section{Saint Venant Transfer Functions} \label{appendix transfer functions}
Linearizing (\ref{StVenant1})-(\ref{StVenant2}) around $(A_0,Q_0)$ by Taylor series, neglecting terms that are quadratic or higher order in $\tilde{A}$ and $\tilde{Q}$, gives a linear PDE of the form
\begin{equation} 
\frac{\partial }{\partial t}  \AQ + B(x) \frac{\partial }{\partial x} \AQ = C(x) \AQ,  \label{StVenant linearized}
\end{equation}
where $B(x)$ and $C(x)$ depend on the pre-computed steady state water profile. Taking the Laplace transform of (\ref{StVenant linearized}) gives a linear ODE for  $\hat{A}$ and $\hat{Q}$ of the form
\begin{equation}
\frac{\partial}{\partial x}\AQs = \Phi(x,s) \AQs, \label{AQs ODE}
\end{equation} 
which leads to a relation of the form
\begin{equation}
\AQout = \Psi(s) \AQin,  \label{Psi}
\end{equation}
where the transition matrix $\Psi(s) $ can be obtained by  solving (\ref{AQs ODE}) numerically. Finally, (\ref{Psi}) can be solved for $\hat{A}(L,s)$. Dividing $\hat{A}(L,s)$ by the equilibrium channel top width, $w(\hrref)$, gives the transfer functions for $\hds$ as in (\ref{transfer function pool i}). 

\section*{\blue Acknowledgements}
\mc{Many important discussions with Dr Adair Lang and Dr Yuping Li at Rubicon Water are gratefully acknowledged. Their assistance with the field trial in particular was indispensable. Thanks also to Goulburn-Murray Water (GMW) for accommodating the trial and consenting to use of the data.}

%\bibliographystyle{IEEEtran}
%\bibliography{IEEEabrv,references}

%\IEEEtriggeratref{22}


%\vfill
%\newpage

\begin{IEEEbiographynophoto}{Timm Strecker} received the M.Sc. degree from the University of Stuttgart, Stuttgart, Germany, in 2015, and the Ph.D. degree in engineering cybernetics from the Norwegian University of Science and Technology (NTNU), Trondheim, Norway, in 2018.

He is currently a Research Fellow with the University of
Melbourne, Melbourne, VIC, Australia. His research interests include modeling and control of distributed parameter systems, and applications in water distribution and drilling.
\end{IEEEbiographynophoto}

% \newpage

%\vspace*{-5in}

\begin{IEEEbiographynophoto}{Michael Cantoni} (Member IEEE) received
  the PhD degree from the University of Cambridge, Cambridge, England,
  in 1998, and the Bachelor of Science (applied mathematics) and
  Bachelor of Engineering (Hons. I, electrical) degrees from the
  University of Western Australia, Crawley, WA, Australia, both in
  1995.

  He is currently a Professor in the Department of Electrical and
  Electronic Engineering at the University of Melbourne, Parkville,
  Australia. From 1998--2000, he was a Research Associate in the
  Department of Engineering, and Title A Fellow at St.~John's College,
  at the University of Cambridge. He held a visiting academic position
  at Imperial College London, England, 2019--20.

  Dr. Cantoni received the IEEE Control Systems Technology Award in
  2014, for the development and implementation of controls for
  irrigation channels and water management with an industry
  partner. He served as Associate Editor for Automatica (2015-2018),
  Systems and Control Letters (2007-2013), and IET Control Theory and
  Applications (2007-2009), and Section Editor for the Springer
  Reference Works Encyclopedia of Systems and Control (2014,
  2019). His research interests include robust and optimal control,
  networked systems, and applications in water and power distribution.
\end{IEEEbiographynophoto}

\end{document}